\documentclass[pdflatex,sn-mathphys-num]{sn-jnl}

\usepackage{graphicx}%
\usepackage{multirow}%
\usepackage{amsmath,amssymb,amsfonts}%
\usepackage{amsthm}%
\usepackage{mathrsfs}%
\usepackage[title]{appendix}%
\usepackage{xcolor}%
\usepackage{textcomp}%
\usepackage{manyfoot}%
\usepackage{booktabs}%
\usepackage{algorithm}%
\usepackage{algorithmicx}%
\usepackage{algpseudocode}%
\usepackage{physics}
\usepackage{listings}%
\usepackage{xcolor}
\usepackage{hyperref}
\usepackage{cleveref}
\usepackage{soul}
\usepackage{quantikz}
\usepackage{indentfirst}
\usepackage[normalem]{ulem}
\usepackage{titles}
\hypersetup{
    colorlinks=true,
    linkcolor=blue,
    urlcolor=blue,
    citecolor=blue
}

\newcommand\redout{\bgroup\markoverwith
{\textcolor{red}{\rule[.5ex]{2pt}{1pt}}}\ULon}

\theoremstyle{thmstyleone}%
\theoremstyle{thmstyletwo}%

\theoremstyle{thmstylethree}%

\begin{document}

%\title[Article Title]{A review of quantum computing and a quantum-inspired methods applied to computational fluid dynamics}

\title[Article Title]{A review of quantum machine learning and quantum-inspired applied methods to computational fluid dynamics}


\author*[1]{\fnm{Cesar A.} \sur{Amaral}}\email{c.amaral@posgrad.ufsc.br}
\equalcont{These authors contributed equally to this work.}

\author*[1]{\fnm{Vinícius L.} \sur{Oliveira}}\email{vinicius.luz.oliveira@posgrad.ufsc.br}
\equalcont{These authors contributed equally to this work.}

\author[2]{\fnm{Juan P. L. C.} \sur{Salazar}}\email{juan.salazar@ufsc.br}

\author[1]{\fnm{Eduardo I.} \sur{Duzzioni}}\email{eduardo.duzzioni@ufsc.br}

\affil*[1]{\orgdiv{Departmento de Física}, \orgname{Universidade Federal de Santa Catarina}, \orgaddress{ \city{Florianópolis}, \postcode{88040-900}, \state{SC}, \country{Brazil}}}

\affil[2]{\orgdiv{Engenharia Aeroespacial}, \orgname{Universidade Federal de Santa Catarina}, \orgaddress{\city{Joinville}, \postcode{89219-600}, \state{SC}, \country{Brazil}}}


\abstract{Computational Fluid Dynamics (CFD) is central to science and engineering, but faces severe scalability challenges, especially in high-dimensional, multiscale, and turbulent regimes. Traditional numerical methods often become prohibitively expensive under these conditions. Quantum computing and quantum-inspired methods have been investigated as promising alternatives. This review surveys advances at the intersection of quantum computing, quantum algorithms, machine learning, and tensor network techniques for CFD.  We discuss the use of Variational Quantum Algorithms as hybrid quantum-classical solvers for PDEs, emphasizing their ability to incorporate nonlinearities through Quantum Nonlinear Processing Units. We further review Quantum Neural Networks and Quantum Physics-Informed Neural Networks, which extend classical machine learning frameworks to quantum hardware and have shown advantages in parameter efficiency and solution accuracy for certain CFD benchmarks. Beyond quantum computing, we examine tensor network methods, originally developed for quantum many-body systems and now adapted to CFD as efficient high-dimensional compression and solver tools. Reported studies include several orders of magnitude reductions in memory and runtime while preserving accuracy. Together, these approaches highlight quantum and quantum-inspired strategies that may enable more efficient CFD solvers. This review closes with perspectives: quantum CFD remains out of reach in the NISQ era, but quantum-inspired tensor networks already show practical benefits, with hybrid approaches offering the most promising near-term strategy. We emphasize that, in addition to being a review, this is also an introductory material to the topics covered.}

%%================================%%
%% Sample for structured abstract %%
%%================================%%




\maketitle

\tableofcontents
\section{Introduction}\label{sec1}

Computational Fluid Dynamics (CFD) has become a central discipline within the broader field of fluid mechanics, with its origins strongly linked to the development of numerical weather prediction in meteorology and to the analysis of aerodynamic flows in aerospace engineering \cite{charney1955primitive, lomax1976fundamentals, Bhattacharyya21} . From these beginnings, CFD has grown into a mature field that transcends its early applications and now constitutes an essential component of nearly all branches of engineering. Its utility is evident across diverse areas, including automotive and aeronautical aerodynamics, turbomachinery design, civil and environmental engineering, and biomedical engineering \cite{Ferziger2012,taebi2024computational, anderson1995cfd}. Beyond traditional engineering domains, CFD is also indispensable in applied physics and astrophysics, where it enables the study of complex plasma flows, stellar dynamics, and accretion processes~\cite{leveque2002finite, stone_et_al_1992, Nishio_et_al_2025}.

The importance of CFD derives from its ability to provide detailed predictions of fluid motion, heat and mass transfer, chemical reactions, and multiphase phenomena in systems where experimental investigation may be prohibitively expensive, dangerous, or even physically infeasible. In particular, CFD plays a critical role in addressing flows governed by the Navier–Stokes equations and related models, which can be expressed in closed mathematical form but rarely admit analytical solutions except in highly simplified cases \cite{batchelor1967fluid, Pope2000}. This limitation is most evident in the study of turbulence, where the inherent nonlinearity and multiscale character of the governing equations make analytical approaches intractable. In such contexts, CFD provides a vital bridge between theoretical analysis and experimental observation.

Advancements in CFD, whether through more accurate discretization schemes, improved turbulence models, or the integration of high-performance computing architectures, have a direct impact on technological innovation. Improvements in CFD capabilities translate into safer and more efficient aircraft, optimized propulsion and combustion systems, enhanced prediction and mitigation of environmental hazards, and better understanding of geophysical and astrophysical flows~\cite{hirsch2007numerical, spalart2009des}. Thus, CFD is not merely a numerical tool but a driver of scientific discovery and industrial competitiveness. Its continued development will remain fundamental to tackling emerging challenges in energy, transportation, climate, and beyond.

Over the past decades, increasingly robust computational strategies have been developed to approximate the governing equations of fluid motion. Among the classical approaches, the finite volume method (FVM), finite element method (FEM), and finite difference method (FDM) remain the most widely applied, with the FVM having established itself as the dominant framework in engineering practice~\cite{Ferziger2012}. Numerous variants and hybridizations of these methods, along with alternative discretization schemes, have been proposed to enhance accuracy and efficiency. Despite these advances, CFD continues to face fundamental challenges in scalability, particularly in the simulation of multiscale, nonlinear phenomena such as turbulence, where coherent structures interact across a broad spectrum of spatial and temporal scales~\cite{Pope2000}. The core difficulties arise from the immense number of degrees of freedom required to resolve fine discretizations of space and time, compounded by the complexity of solving coupled systems with multiple interacting physical variables. These constraints underscore the pressing need not only for improved mathematical formulations and advanced numerical algorithms, but also for the exploration of novel computational paradigms capable of addressing the intrinsic high dimensionality and complexity of CFD.

An example of such a paradigm is quantum computing. Owing to the fact that the Hilbert space associated with a quantum system grows exponentially with the number of qubits, quantum computers can, in principle, represent exponentially large vectors at the cost of a modest number of quantum bits. Entanglement also provides a natural mechanism for encoding nontrivial correlations between physical variables, as is the case in fluid turbulence~\cite{gourianov_quantum_2022}.

Inspired by classical discretization techniques, a natural strategy is to use quantum algorithms to solve the large linear systems that arise from discretized PDEs~\cite{montanaro2016quantum, pollachini2021hybrid}. This approach is attractive because linear solvers typically dominate computational cost. The HHL algorithm, for example, offers an exponential speedup in specific circumstances~\cite{harrow2009quantum}. In fact, HHL has been applied in principle to PDEs in various contexts, including electromagnetism, thermodynamics, and acoustics~\cite{clader2013preconditioned, cao2013quantum, costa2019quantum, linden2022quantum}. However, practical implementation of HHL lies beyond the reach of current Noisy Intermediate-Scale Quantum (NISQ) devices and may remain impractical even with fault-tolerant quantum computers~\cite{scherer2017concrete}. This motivates the exploration of alternative strategies~\cite{Jaksch2023, Bosco2024}, most notably Variational Quantum Algorithms (VQAs)~\cite{cerezo_variational_2021} and Quantum Machine Learning (QML) approaches~\cite{schuld_circuit-centric_2020}.

The central idea of a VQA is to use a parameterized quantum circuit to prepare a quantum state, whose measurement defines a cost function, such as the expected value of an observable or the fidelity with a specific target state. This cost is optimized by a classical routine (e.g., gradient descent) that updates the circuit parameters. By leveraging superposition and entanglement, VQAs can explore complex solution landscapes with relatively few qubits, making them one of the most practical strategies for the NISQ era. QML, in turn, focuses on using quantum circuits to learn patterns from data~\cite{biamonte_quantum_2017, du_quantum_2025}. In gate-based quantum computing, this often reduces to training parameterized quantum circuits (PQCs). In the context of PDEs, the most relevant class of QML models is Quantum Neural Networks (QNNs), which serve as quantum analogs of classical neural networks. Because QNNs are usually optimized via variational training, the terms QML and VQA are often used interchangeably, though strictly speaking, ML implies data-driven function approximation~\cite{roberts_principles_2022}. A common QML strategy is therefore to train PQCs using VQA techniques, defining the cost function in terms of a dataset~\cite{mangini_quantum_2021}.

The use of machine learning in CFD is well established \cite{wang_recent_2024, li_fourier_2021, kochkov_machine_2021, jin_nsfnets_2021}, but the use of QML is relatively new. In particular, Quantum Physics-Informed Neural Networks (QPINNs) have emerged as a promising framework. Classical PINNs incorporate physical laws directly into their loss function~\cite{raissi_physics-informed_2019}, reducing dependence on data and improving interpretability. QPINNs retain these advantages while also including the benefits of using parameterized circuits to encode the solution. Recent results demonstrate that QPINNs can achieve better approximations than classical PINNs in some cases~\cite{trahan_quantum_2024, farea_qcpinn_2025}.

Despite these prospects, both VQAs and QML face practical challenges, including quantum noise and barren plateaus, which complicate training, especially in deeper circuits, and hinder scalability~\cite{cerezo_variational_2021, Larocca2025, Qi2023}. This has spurred interest in other methods for addressing high dimensionality that are already more accessible. For example, Au-Yegun et al. recently applied quantum walks to solve the one-dimensional advection equation in a CFD context~\cite{Au2025}. Another powerful alternative is Tensor Networks (TNs) \cite{Schollwock2011}. Originally developed for Quantum Many-Body Systems, TNs also tackle the curse of dimensionality by representing states and operators as networks of low-rank tensors. These tensors carry physical indices as well as virtual bond indices, with the bond dimension controlling the retained correlations. By truncating to manageable bond dimensions, one obtains compact representations that capture dominant correlations while drastically reducing the computational resources needed.

This strategy has been successfully extended to classical high-dimensional problems, including CFD. In these applications, physical fields are mapped onto tensor structures, enabling efficient simulations at reduced cost, especially for high-resolution domains. Reported results are striking: Nikita et al.~\cite{Nikita2025} achieved a $10^3$ runtime reduction and $10^6$ memory reduction compared to dense solvers, while Kiffner et al.~\cite{Kiffer2023} demonstrated $97\%$ compression and up to $17\times$ speed-up. Similar gains have been reported in other works comparing TN-based solvers with FEM, FVM, and FDM~\cite{Pisoni2025, Danis2025, Kornev2023, Holscher2025, Danis2024, Dubois2025, Walton2025, Peddinti2024, Vanhulst2025}. 

The convergence of quantum computing and quantum-inspired methods is fostering a promising set of tools for the next generation of fluid dynamics solvers. This review aims to provide an accessible entry point into the field, where we also illustrate the methods used with the problem of the 1D viscous Burgers' equation and provide codes for each method on our GitHub\footnote{Available at: \url{https://github.com/cammaral/quantum_and_cfd/}.}. In Section~\ref{vqa}, we examine VQAs for solving typical PDEs in fluid dynamics, highlighting how nonlinear problems can be addressed through a Quantum Nonlinear Processing Unit~\cite{lubasch_variational_2020}. Section~\ref{QNNs_and_QPINN} explores the use of QNNs, with particular emphasis on recent efforts to extend PINNs to the quantum framework for CFD applications. Section~\ref{quantum_insp_cesar} introduces tensor network–based solvers, illustrating how quantum-inspired methods can be leveraged in CFD.
Finally, Section~\ref{sec:perspectives} outlines open challenges and future directions.

\section{Variational Quantum Algorithms}
\label{vqa}
Variational Quantum Algorithms (VQAs) are hybrid optimization algorithms designed to exploit the strengths of both quantum and classical computing. They are among the most promising approaches for achieving quantum advantage within the constraints of the current NISQ era~\cite{cerezo_variational_2021, mangini_variational_2023}. The core idea is to create a parametrized quantum state on a quantum processor while delegating the optimization of parameters to a classical computer. In this way, VQAs circumvent some of the hardware limitations of noisy quantum devices, such as noise mitigation via shallow circuits and sample-based measurements~\cite{cerezo_variational_2021}, while leveraging well-established tools from nonconvex optimization.

For a given task, the main components of a VQA are:
\begin{itemize}
    \item \textbf{Parameterized quantum circuit (ansatz):} An $N$-qubit unitary circuit $U(\boldsymbol{\theta})$ with tunable parameters $\boldsymbol{\theta} \in \mathbb{R}^K$, to prepare the state
    \[
    \ket{\psi(\boldsymbol{\theta})} = U(\boldsymbol{\theta})\ket{0}^{\otimes N}.
    \]
    The architecture of the circuit (depth, gate set, connectivity, etc.) is chosen at the start of the algorithm and is referred to as the \emph{ansatz}.
    \item \textbf{Cost function:} A trainable function $C(\boldsymbol{\theta}): \mathbb{R}^K \to \mathbb{R}$ that encodes the task and quantifies the performance of the circuit. By construction, minimizing $C(\boldsymbol{\theta})$ corresponds to solving the problem of interest.
    \item \textbf{Observables:} A set of self-adjoint operators $\{A_k\}$ whose expectation values define the cost function.
\end{itemize}

The cost function is expressed in terms of the quantum state and observables as
\begin{equation}
C(\boldsymbol{\theta}) = C(U(\boldsymbol{\theta}); \{ A_{k} \})\,\,.
\end{equation}

\begin{figure}
    \centering
    \includegraphics[width=0.7\linewidth]{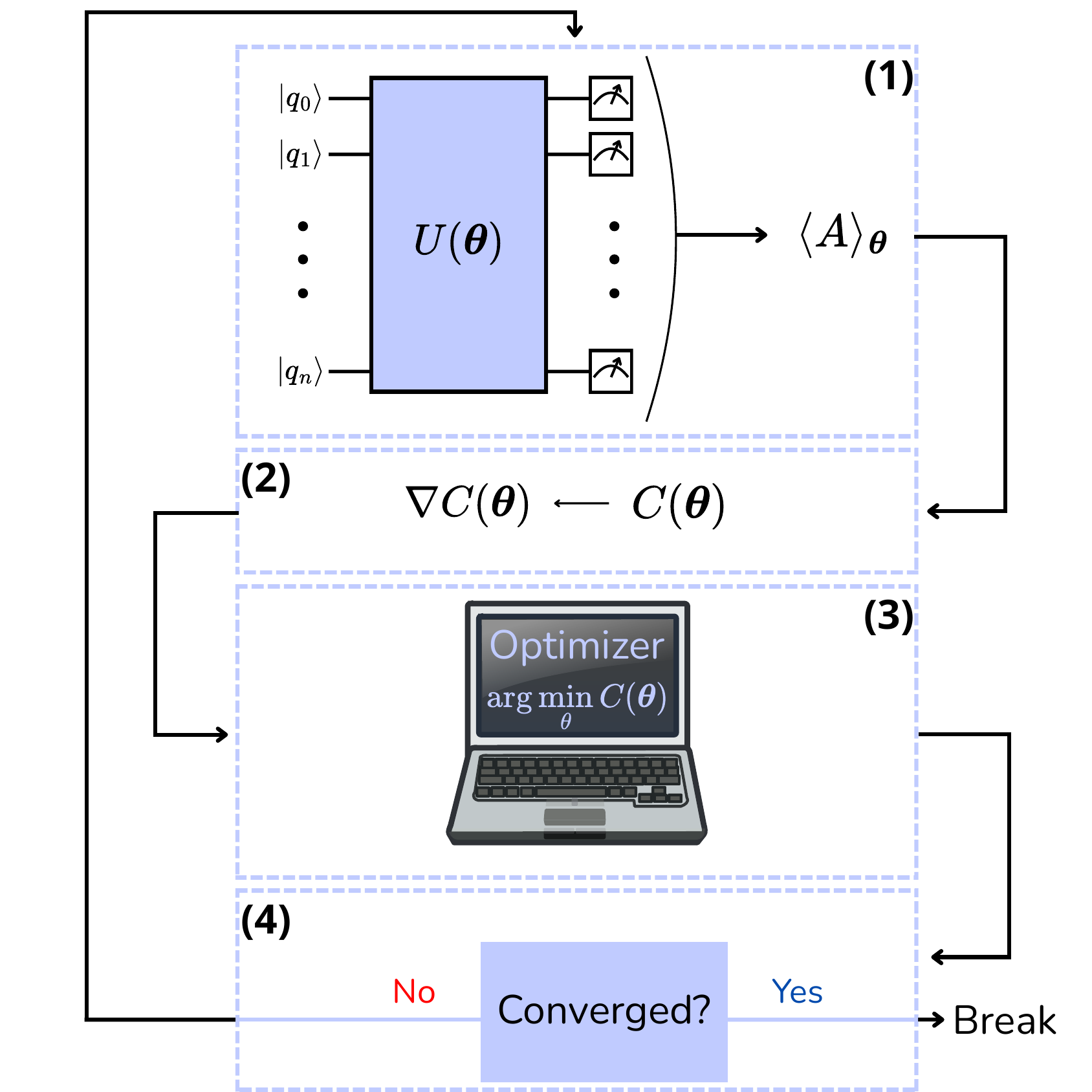}
    \caption{Schematic of the hybrid loop in a VQA. (1) The parameterized circuit $U(\boldsymbol{\theta})$ is executed and observables are measured. (2) The cost function (and if necessary, gradients) are computed. (3) A classical optimizer updates the parameters, which are (4) fed back into the quantum circuit. These steps are repeated for a prescribed number of iterations or until a convergence criterion is met.}
    \label{vqa_diagram}
\end{figure}

The algorithm proceeds iteratively in the following loop:
\begin{enumerate}
    \item \textbf{Quantum evaluation:} The parameterized circuit $U(\boldsymbol{\theta})$, as shown in figure \ref{vqa_diagram}, is executed, and repeated measurements are performed to estimate expectation values
    \[
    f_{\boldsymbol{\theta}} := \bra{\psi(\boldsymbol{\theta})}A_k\ket{\psi(\boldsymbol{\theta})}\,\,.
    \]
    \item \textbf{Cost and gradients:} The cost function is evaluated, and its gradient is estimated. Methods for gradient evaluation in parameterized circuits are discussed in Section~\ref{gradients}.
    \item \textbf{Classical optimization\footnote{There are alternative optimization methods that do not use gradients, such as quantum annealing~\cite{streif_comparison_2019, verdon_quantum_2019} or usage of Bayesian optimization methods~\cite{iannelli_noisy_2021}. VQAs started out using a direct search method~\cite{peruzzo_variational_2014}. However, in this work, we will focus only on gradient methods.}} A classical optimizer updates the parameters,
    \begin{equation}
    \boldsymbol{\theta}^{(t+1)} = \boldsymbol{\theta}^{(t)} - \eta \, \nabla_{\boldsymbol{\theta}} C(\boldsymbol{\theta}^{(t)})\,\,,
    \end{equation}
    where $t$ denotes the iteration step and $0 < \eta \ll 1$ is the learning rate, controlling the step size and ensuring stable convergence.
    \item \textbf{Iteration:} Steps (1)--(3) are repeated for a prescribed number of iterations or until a convergence criterion is met.
\end{enumerate}

Figure~\ref{vqa_diagram} illustrates the hybrid loop of a VQA according to the steps above.

\subsection{Cost functions and gradient methods} \label{gradients}

We now expand on cost functions and their optimization within VQAs, as the construction of a cost function is the interface between the optimization task and the CFD problem. Recall that a cost function must be \emph{trainable}, meaning that it possesses a minimum which corresponds to the solution of the problem. Formally, the optimal parameter configuration $\boldsymbol{\theta}^*$ is defined as
\begin{equation}
\boldsymbol{\theta}^* = \text{arg } \min_{\boldsymbol{\theta}}C(\boldsymbol{\theta})\,\,.
\end{equation}

There are many different strategies for defining a cost function for a VQA, such as fidelity-based costs~\cite{cerezo_fidelity_2020, romero_quantum_2017}, divergence-based costs~\cite{dallaire-demers_quantum_2018} or ``information-theoretical based'' using quantum resources such as purity~\cite{tan_variational_2021}. As stated in section \ref{vqa}, a common approach that will be adopted in this work is to express cost functions in terms of simple expectation values of observables. This provides two main advantages. First, it simplifies readout: the optimization is performed over classical parameters $\boldsymbol{\theta}$, so the entire quantum state does not need to be reconstructed. Second, it leverages the exponentially large Hilbert space of the quantum device while still producing a scalar quantity suitable for classical optimization. Concretely, once the circuit is executed, measurements of an observable $A$ are performed. With $S$ shots, the empirical expectation value is estimated as

\begin{equation}
\hat{f}_{\boldsymbol{\theta}} = \frac{1}{S}\sum_{k=1}^Sa_{k}\,\,, 
\end{equation}
where $a_k$ is the $k$-th measurement outcome and $\hat{f}_{\boldsymbol{\theta}}$ is the estimator for $f_{\boldsymbol{\theta}}$. The statistical error $\epsilon$ scales as $1/\sqrt{S}$. Although the output is inherently stochastic, posing challenges for gradient-based optimization, this issue can be mitigated with suitable gradient estimation techniques~\cite{mari_estimating_2021}.

A further advantage of this hybrid framework is that optimization parameters remain classical variables, requiring no additional qubits for storage and thus avoiding memory overhead on quantum hardware.

In classical machine learning, gradient descent is the standard approach for optimizing models such as feed-forward neural networks, where the cost function is often the mean squared error relative to target data~\cite{roberts_principles_2022}. Gradients are typically computed using automatic differentiation (autodiff), which combines the chain rule with numerical evaluation. Formally, for a cost function $C(\boldsymbol{\theta}) = C(f_{\boldsymbol{\theta}})$, the derivative with respect to the $j$-th parameter is
\begin{equation}
\frac{\partial C}{\partial \theta_{j}} = \frac{\partial C}{\partial f_{\boldsymbol{\theta}}} \frac{\partial f_{\boldsymbol{\theta}}}{\partial\theta_{j}}\,\,.
\end{equation}

In VQAs, the expectation value $
f_{\boldsymbol{\theta}} = \bra{\psi(\boldsymbol{\theta})} A \ket{\psi(\boldsymbol{\theta})} $
is itself a random variable estimated from the measurements, and its derivative $\partial_{\theta_j} f_{\boldsymbol{\theta}}$ is not straightforward. Naïve finite-difference approximations are impractical: small step sizes require extremely precise expectation values, leading to prohibitively many measurements~\cite{guerreschi_practical_2017}. Moreover, differentiating the circuit directly involves derivatives of unitary gates, which may not remain unitary.

To overcome these issues, the most widely used approach is the \emph{parameter-shift rule}~\cite{schuld_effect_2021}. It states that for certain parameterized gates, the gradient of an expectation value can be written as

\begin{equation}
\frac{\partial{f_{\boldsymbol{\theta}}}}{\partial \theta_{j}} = \sum_{i} c_{i}f_{\boldsymbol{\theta} + s_{i}}\,\,,
\end{equation}
where $\{ c_{i} \}$ and $s_{i}$ are real values. A parametrized quantum gate can be expressed in exponential form as $
W(\theta_j) = e^{-i\frac{\theta_j}{2}G_j}$, 
with generator $G_j = G_j^\dagger$. For rotation gates, where $G_j^2 = I$ (involutory generators), derivatives can be written~\cite{mari_estimating_2021} as:

\begin{equation}
    \label{ps_1}
   \frac{\partial W(\theta_{j})}{\partial \theta_{j}}  = \frac{W(\theta_{j} + s) - W(\theta_{j}-s)}{2 \sin(s)}\,\,, 
\end{equation}
valid for any $s$ within the periodic domain of the gate. This can be used to calculate the $j$-th entry of the gradient $g_j(\boldsymbol{\theta})$. Considering $\mathbf{e}_j$ the $j$-th basis vector of $\mathbb{R}^K$, choosing $s = \frac{\pi}{2}$, and applying Eq. \cref{ps_1}, the following expression holds for the analytical gradient:
\begin{equation}
    \label{ps_2}
g_{j}(\boldsymbol{\theta}) = \frac{f_{\boldsymbol{\theta}}\left( \boldsymbol{\theta}+\frac{\pi}{2} \mathbf{e}_{j} \right)-f_{\boldsymbol{\theta}}\left( \boldsymbol{\theta}-\frac{\pi}{2}\mathbf{e_{j}} \right)}{2}\,\,.
\end{equation}
Higher-order derivatives can be obtained by iteratively applying the parameter-shift rule to Eq.~\eqref{ps_1}, with the number of terms doubling at each order
\cite{mari_estimating_2021}.

For $m$ parameters and $S$ shots, the computational complexity of gradient evaluation via parameter-shift is $\mathcal{O}(2 m S)$. Although this scaling appears less efficient than classical autodiff, the required number of shots can remain modest, and convergence to a minimum is guaranteed under suitable conditions~\cite{sweke_stochastic_2020}. For gates with non-involutory generators, alternative decompositions (e.g., into products of rotation gates) or generalizations of the parameter-shift identity can be employed~\cite{crooks_gradients_2019}.

The parameter-shift rule provides a practical and widely adopted method for gradient estimation in VQAs. While many other techniques exist, their applicability depends on the structure of the quantum circuit and the problem at hand. By combining low measurement overhead with classical optimization, VQAs offer a powerful framework for addressing high-dimensional problems such as those encountered in CFD.

\subsection{VQAs for fluid dynamics}
\label{dietter_method}

Whether iterative linear solvers or variational optimization yield the greater quantum speedup remains an open problem-dependent question. Nevertheless, the variational approach offers a flexible framework that is especially appealing for the inherently multiscale dynamics of CFD. The methods in~\cite{lubasch_variational_2020, jaksch_variational_2023, Leong2022, Leong2023, Leong2024} cast the PDE solution as the minimizer of a suitably designed cost functional, optimized over a trial function (ansatz) implemented by a parameterized quantum circuit:

\begin{enumerate}
    \item  For a given PDE, specify a trial function $f_{\boldsymbol{\theta}}$ (or a set of fields), with tunable parameters $\boldsymbol{\theta} \in \mathbb{R}^K$. The ansatz should be expressive enough to approximate the target solution while remaining trainable.

    \item Discretize the spatial (and, if needed, temporal) domain to obtain a set of collocation points and boundary/initial samples. Realize $f_{\boldsymbol{\theta}}$ on a quantum device by applying a parameterized circuit $U(\boldsymbol{\theta})$ to $\ket{\mathbf{0}}:= \ket{0}^{\otimes N}$ and reading out appropriate observables so that $\langle A \rangle_{\psi(\theta)}$ returns the required field values and/or their components at the sampled locations.

    \item Construct a cost $C(\boldsymbol{\theta})$ whose minimum corresponds (up to discretization error) to a PDE solution. Typical terms include the mean-squared PDE residual on interior collocation points, penalties enforcing boundary/initial conditions, and optional regularization.

    \item Implement the (possibly nonlinear) components of $C(\boldsymbol{\theta})$ within the circuit using an additional processing layer—termed the \emph{Quantum Nonlinear Processing Unit} (QNPU) in~\cite{lubasch_variational_2020}. The QNPU uses ancilla-assisted constructions and controlled operations to evaluate nonlinear functionals (e.g., products, norms) from measurement statistics without reconstructing the full state.

    \item Measure an ancilla qubit (and the relevant system registers) to estimate the cost and, via parameter-shift or related rules, its gradients. A classical optimizer then updates $\boldsymbol{\theta}$; the process is iterated until convergence or a stopping criterion is met.
\end{enumerate}

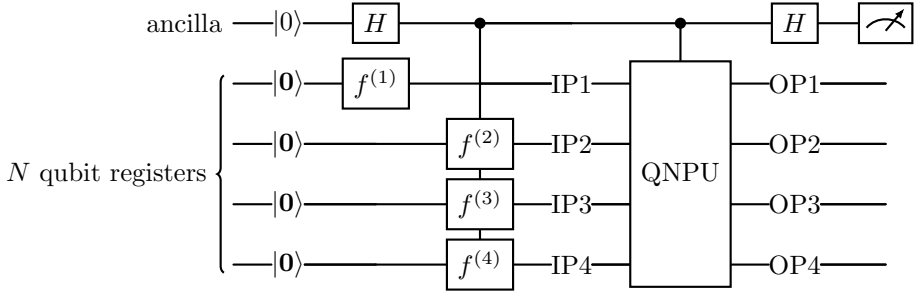
\begin{figure}
\begin{quantikz}[row sep={0.8cm,between origins}, column sep=0.5cm]
\lstick[wires=1]{ancilla} & \ket{0} & \gate{H} & \ctrl{4} & \qw & \ctrl{1} &  \gate{H} & \meter{} \\
\lstick[wires=4]{$N$ qubit registers} & \ket{\mathbf{0}} & \gate{f^{(1)}} & & \qw \text{IP1}& \gate[4][1cm]{\text{QNPU}} & \text{OP1} & \qw\\
 & \ket{\mathbf{0}} & \qw & \gate{f^{(2)}} & \text{IP2} &\qw & \text{OP2}& \qw \\
 & \ket{\mathbf{0}} & \qw & \gate{f^{(3)}} & \text{IP3}  & \qw & \text{OP3}& \qw \\
 & \ket{\mathbf{0}} & \qw & \gate{f^{(4)}} & \text{IP4} & \qw  & \text{OP4}& \qw
\end{quantikz}
    \caption{Variational circuit for the method in Section~\ref{dietter_method}. Each bold wire $\ket{\mathbf{0}}$ denotes a register (here shown as four registers for clarity). The circuit prepares multiple functions via gates $f^{(j)}$ (distinct dependent variables or replicated copies), which are fed as inputs (IP) to the controlled QNPU block that evaluates the nonlinear cost components and produces outputs (OP). The ancilla measurement yields the cost (and, with parameter-shift, its gradients), which are used by a classical optimizer to update the parameters. These steps are repeated until a convergence criterion is met.}
    \label{circuit_dieter}
\end{figure}

The circuit and general workflow are summarized in Figure \ref{circuit_dieter}.

\subsubsection{Encoding and trial functions}
\label{encoding_on_vqa}
Consider a discretization $A \subset \mathbb{R}^n$ of the PDE domain, where $A$ is an exponentially fine $n$-dimensional grid with $2^N$ points along each axis, resulting in a total of $2^{N \cdot n}$ grid points. Our goal is to approximate the solution $u: A \to \mathbb{R}^r$, which in general is a vector field. To build intuition, we first restrict to the simple case $n=r=1$ (as in the Burgers equation example) and later generalize to arbitrary vector fields on higher-dimensional grids.

Nonlinear PDEs are notoriously chaotic: small variations in initial or boundary conditions may lead to drastically different solutions, as in turbulence. Nevertheless, one can begin with an initial guess $f$, the \emph{ansatz}, which serves as a starting point for the variational optimization. Through iterative updates, the trial function is refined until $f \approx u$.

To use this ansatz within a quantum circuit, it must be encoded in the quantum state of $N$ qubits. Among the many read-in strategies available~\cite{schuld_effect_2021, rath_quantum_2024}, amplitude encoding~\cite{rath_quantum_2024} is a natural choice. Let $\mathbf{f} := f(\mathbf{r}) \in \mathbb{R}^{2^N}$ denote the discretized trial function evaluated at all grid points $\mathbf{r}$. Normalizing $\mathbf{f}$ yields the quantum state
\begin{equation}
\ket{f}  = \sum_{j=0}^{2^N-1}f_{j}\ket{j} , \quad j \in \{ 0, 1 \}^N\,\,,
\end{equation}
where $f_{j} = \mathbf{f_{j}}/\lvert \lvert \mathbf{f} \rvert \rvert$. A block of parameterized quantum gates is set with variational parameters $\boldsymbol{\theta}$ such that,  
\begin{equation}
U(\boldsymbol{\theta})\ket{\mathbf{0}} = \ket{f}\,\,.
\end{equation}
and the parameters $\boldsymbol{\theta}$ are optimized via a cost function so that $\ket{f}$ encodes a function closer to the PDE solution.
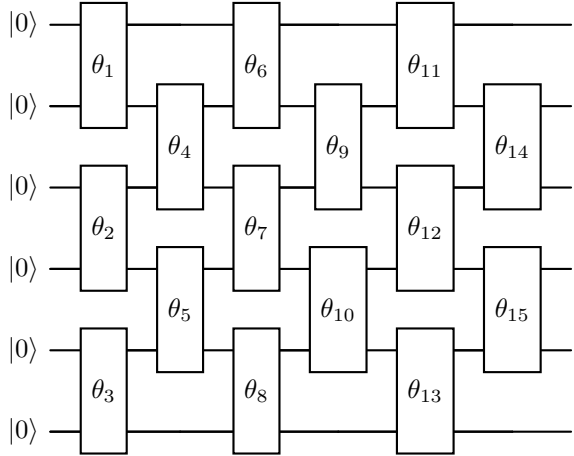
\begin{figure}
    \centering
    \begin{quantikz}[column sep=0.4cm]
  \lstick{$\ket{0}$} & \gate[2]{\theta_1}  & \qw                    & \gate[2]{\theta_6}  & \qw                     & \gate[2]{\theta_{11}} & \qw                     &  \\
  \lstick{$\ket{0}$} & \qw                    & \gate[2]{\theta_4}  & \qw                    & \gate[2]{\theta_9}   & \qw                      & \gate[2]{\theta_{14}} &  \\
  \lstick{$\ket{0}$} & \gate[2]{\theta_2}  & \qw                    & \gate[2]{\theta_7}  & \qw                     & \gate[2]{\theta_{12}} & \qw                      &  \\
  \lstick{$\ket{0}$} & \qw                    & \gate[2]{\theta_5}  & \qw                    & \gate[2]{\theta_{10}}  & \qw                      & \gate[2]{\theta_{15}} &  \\
  \lstick{$\ket{0}$} & \gate[2]{\theta_3}  & \qw                    & \gate[2]{\theta_8}  & \qw                     & \gate[2]{\theta_{13}} & \qw                      &  \\
  \lstick{$\ket{0}$} & \qw                    & \qw                    & \qw                    & \qw                     & \qw                      & \qw                      & 
\end{quantikz}
    \caption{Example of ansatz for the variational circuit for $N=6$ qubits, following a Matrix Product State ansatz as described in \cite{lubasch_variational_2020}.}
    \label{variational_ansatz}
\end{figure}

For vector-valued solutions ($r > 1$), additional trial functions are introduced. For instance, for a 3D velocity field $u = (u_x, u_y, u_z)$, define trial functions $f^{(1)}, f^{(2)}, f^{(3)}: A \to \mathbb{R}$ and optimize them in parallel: $\ket{f} = \ket{f^{(1)}}  \ket{f^{(2)}}  \ket{f^{(3)}}
$. This approach generalizes to any number of dependent variables, limited only by available qubits. The same trial function can also be encoded multiple times to facilitate nonlinear operations in the QNPU.

Although amplitude encoding is conceptually simple, it poses practical challenges, such as state preparation complexity and the normalization constraint, which can distort data. Readout of the entire solution requires exponentially many measurements, which constitutes a bottleneck. Fortunately, often in CFD global quantities are of most interest (e.g., energy norms, total fluxes and forces) or localized features, not the full state, which alleviates this limitation.

In order to better capture the multiscale structure of CFD solutions, an alternative is \emph{multigrid encoding}~\cite{lubasch_multigrid_2018, gourianov_quantum_2022}. Inspired by Matrix Product States (MPS), this method employs a hierarchical Schmidt decomposition from coarse to fine grids, efficiently encoding correlations across scales. The resulting ansatz scales polynomially in the number of qubits, as shown in Fig.~\ref{variational_ansatz}. Details on MPS are given in Section~\ref{quantum_insp_cesar} and in the supplementary material of~\cite{lubasch_variational_2020}. We first illustrate the procedure for the one-dimensional spatial case, and then generalize it.

\begin{figure}
    \centering
    \includegraphics[width=0.7\linewidth]{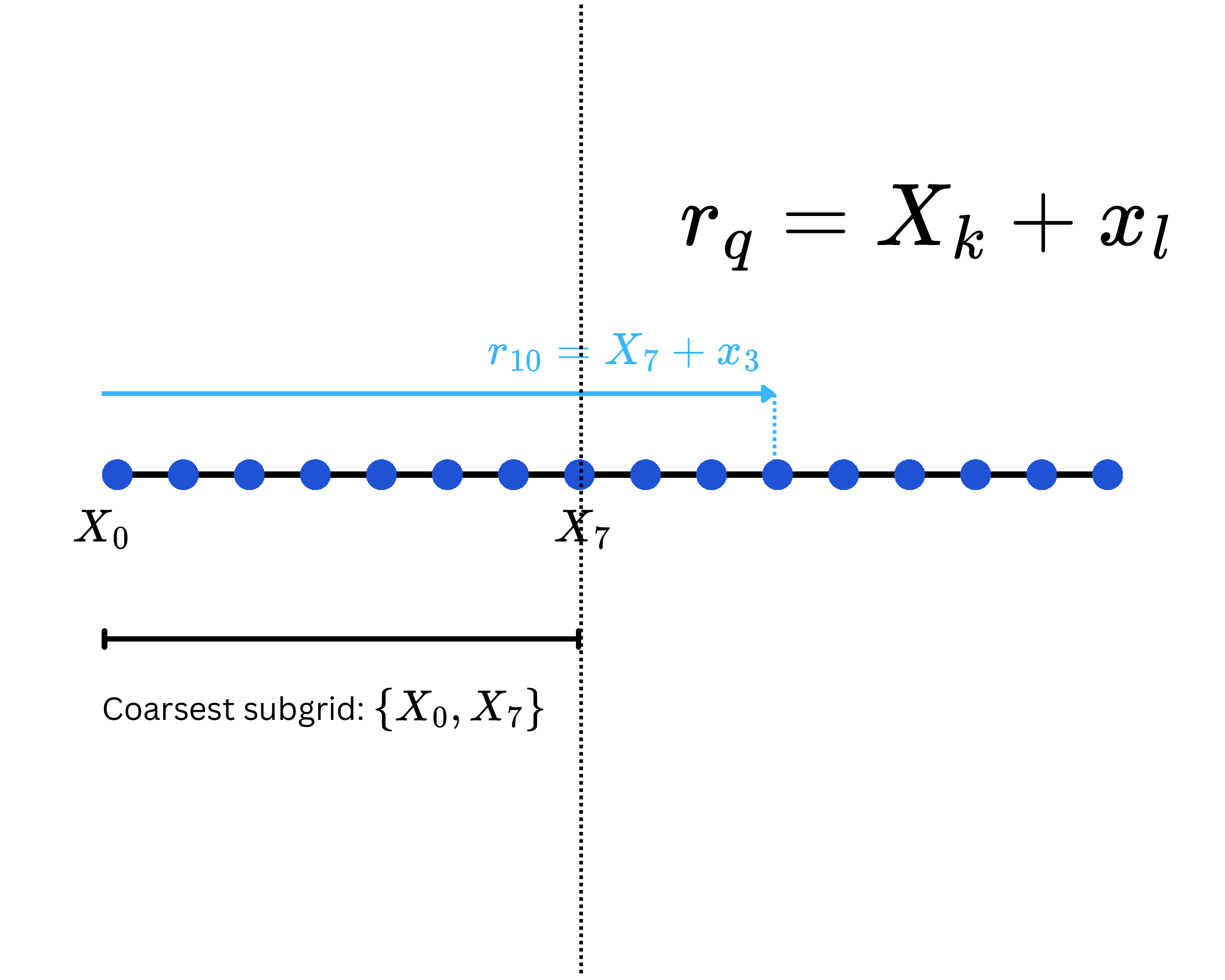}
    \caption{Example of a discretization procedure for multigrid encoding in a one-dimensional problem. Here, $N=4$, with a total of 16 points. The coarsest subgrid is composed of points $\{X_0, X_7\}$. This separation of space is used to define the matrix space utilized to kickstart the consecutive SVDs.}
    \label{illustrating_discretization}
\end{figure}

Consider a one-dimensional spatial domain $[0,L]$ discretized into $2^N$ points, with positions $\mathbf{r}_q = L q / 2^N$ for $q=0,\dots,2^N-1$. Define a bipartition of the grid into a coarse subgrid of $2^m-1$ points $\mathbf{X}_k = Lk/2^m$ ($k=0,\dots,2^m-1$), and for each $\mathbf{X}_k$ a fine subgrid of points $x_l = L l / 2^{N-m}$ ($l=0,\dots,2^{N-m}-1$). Then each position admits the unique decomposition $\mathbf{r}_q = \mathbf{X}_k + \mathbf{x}_l$, and the trial function can be written as $f(\mathbf{r}_q) = f(\mathbf{X}_k + \mathbf{x}_l)$ (Fig.~\ref{illustrating_discretization}).

Take the coarsest sub-grid possible, i.e, a bipartition such that $m=1$. Perform a Schmidt decomposition in order to separate $\mathbf{X}_{k}$ and $\mathbf{x}_l$ terms in the trial function vector $\mathbf{f}$. For that, consider $\mathbb{R}^{2^N} \cong \mathbb{R}^2 \otimes \mathbb{R}^{2^{N-1}}$ and so a SVD on its rectangular matrix representation $f(\mathbf{r}_q) \in \mathbb{R}^{2 \times 2^{N-1}}$ - here rows correspond to steps along the coarse grid, and columns along the fine grid. This gives us
\begin{equation}
f(\mathbf{r}_q) = \sum_{\alpha=1}^2 \lambda_{\alpha}R_{\alpha}(\mathbf{X}_k)f_{\alpha}(\mathbf{x}_l)\,\,,
\label{schmidt1}
\end{equation}
where $\lambda_{\alpha}$ are the positive-valued Schmidt numbers, $R_{\alpha}(\mathbf{X}_k)$ and $f_{\alpha}(\mathbf{x}_l)$ are each orthogonal functions (with respect to different values of $\alpha$) defined uniquely on the coarsest and finer grids, respectively.
The power of this representation is that $R_{\alpha}(\mathbf{X}_k)$ can be written in the first two qubits of the quantum register, capturing a representation of the trial function in this coarse sub-grid such that the state can now be thought of $\ket{\psi} = \sum_{\alpha} \lambda_{\alpha} \ket{R_{\alpha}}\ket{f_{\alpha}}$,  where $\ket{R_{\alpha}}$ is the standard amplitude encoding of $R_{\alpha}(\mathbf{X}_k)$.

The procedure can be done recursively, dividing the two finer grids of points $\mathbf{x}_l$, with $2^{N-1}$, into four grids of $2^{N-2}$ points and performing a Schmidt decomposition in the same fashion as in Eq. \eqref{schmidt1}:
\begin{equation}
f(\mathbf{r_{q}})=\sum_{\alpha=1}^2 \lambda_\alpha R_\alpha\left(\boldsymbol{X}_{k}\right) \sum_{\beta=1}^2 \lambda_{\alpha \beta} R_{\alpha \beta}\left(\boldsymbol{Y}_{k}\right) f_{\alpha \beta}\left(\boldsymbol{y}_{o}\right)\,\,,
\label{schmidt2}
\end{equation}
with $\mathbf{Y}_k+\mathbf{y}_o = \mathbf{x}_l$ and getting the state representation:
\begin{equation}
\ket{\psi}  = \sum_{\alpha =1}^2\ket{R_{\alpha}} \sum_{\beta =1}^2 \lambda_{\alpha \beta} \ket{R_{\alpha \beta}}  \ket{f_{\alpha \beta }}\,\,.
\end{equation}
By covering the entire set of $2^N$ points recursively, each qubit progressively stores a finer resolution. In this way, it is possible to zoom in on regions and parameterize the quality of features in finer representations, tracing out the least significant qubits.

A second advantage is the possibility of truncating the weighted Schmidt values by dropping small values $\lambda_{\alpha}$ while retaining acceptable fidelity to the approximation. Here, the entanglement quantified by the Schmidt values stores correlations of features between length scales, leading to an efficient representation of flows with low correlation between scales by requiring less entanglement (and, therefore, rendering more feasible circuits).

The method readily generalizes to higher-dimensional problems with two- or three-dimensional grids. The only modification concerns the meaning of the ``coarsest grid'': it corresponds to a $2 \times 2$ grid in two dimensions or a $2 \times 2 \times 2$ grid in three dimensions. In other words, the smallest bipartition of the system is characterized by $m=2$ or $m=3$, respectively. This change clearly affects the bond dimension (the smallest dimension of the decomposition of the tensor product space) that depends on $2^m$, giving us, per decomposition, a total of four Schmidt numbers for 2D and eight for 3D, requiring larger matrices.

\subsubsection{Basic structure and cost functions}

To quantify the discrepancy between a trial function $f^{(i)}$ and the true solution $u$, and to guide parameter updates, we recast the dynamics into a cost functional $C(\boldsymbol{\theta}) \equiv C\!\left(f^{(i)}\right)$. In some cases, this is immediate: if the system admits a Lagrangian or Hamiltonian description, one can minimize that functional directly~\cite{lubasch_variational_2020}, or use an action principle including time as an independent variable. However, many fluid PDEs contain dissipative terms that break time-reversal symmetry and lack a simple variational formulation. In such settings, a practical alternative is to adopt a single explicit Euler time step and define a cost that penalizes its residual.

In what follows, we illustrate the procedure using the 1D viscous Burgers equation ~\citep{BURGERS1948171},
\begin{equation}
\label{burgers_eq}
\frac{\partial u(x,t)}{\partial t} = -u(x,t)\frac{\partial u(x,t)}{\partial x} + \nu \frac{\partial^2 u(x,t)}{\partial x^2}\,\,,
\end{equation}
where $\nu$ is the kinematic viscosity and $u(x,t)$ is the velocity field, which depends on the spatial $x$ and temporal $t$ coordinates.

For a time step $\tau$ and a spatial discretization, rearranging the explicit Euler update for Burgers’ equation yields

\begin{equation}
f(x, t + \tau) - [1 + \tau(\nu \Delta - D_{f}\nabla )]f(x, t)= 0\,\,,
\label{euler}
\end{equation}
where $\nabla$ denotes the discrete spatial derivative, $\Delta$ the discrete Laplacian, and $D_f$ the diagonal multiplication operator whose entries are the values of $f(x,t)$ at the grid points. The extension to higher spatial dimensions is immediate by promoting $\nabla$ and $\Delta$ to their $n$-D discrete counterparts.

We can define linear operators $O_{i}$ with respect to each dependent variable $f^{(i)}$ in terms of the linear operators present in the variational form, such as in Eq. \eqref{euler}. For example, 
\begin{equation}
O(t) := \nu \Delta - D_{f}\nabla\,\,.
\label{operator}
\end{equation}
To define the cost function, we take the square of the norm $\lvert \lvert \cdot \rvert \rvert_{2}$ of the variational form. While the cost function will still be in general nonconvex, it will be at least positive-valued. The advantage of taking the $\lvert \lvert \cdot \rvert \rvert_{2}$ norm is the natural way of accessing it in quantum mechanics -  while calculating the norm would imply handling some challenging integrals, we can obtain these values by simply measuring the registers after going through the QNPU.

The class of cost functions $\mathcal{C}: \mathbb{R}^k \to \mathbb{R}$ we can cast doing this have the following form:
\begin{equation}
\mathcal{C} = f^{(1)}\prod_{i=1}^r(O_{i}f^{(i)})\,\,,
\label{cost_family}
\end{equation}
where $r$ is the number of dependent variables we are interested in (e.g, for incompressible 3D Navier-Stokes, these are the three velocity components plus pressure, $r=4$). This is a useful way to modularize the function and translate it to quantum gates.

In the viscous Burgers equation example, there is only one velocity field component (time is not a coordinate) so $r=1$. Therefore, we can define the cost function:
\begin{equation}
C(\boldsymbol{\theta}) = \lvert \lvert f_{\boldsymbol{\theta}}(x, t+\tau)  - (1+\tau O(t))f_{\boldsymbol{\theta}}(x, t)    \rvert  \rvert _{2}^2\,\,.
\label{cost_burgers}
\end{equation}

Dissipative PDEs (e.g., Burgers) do not, in general, preserve the norm of the field, whereas circuit evolution is unitary. To decouple state preparation from norm changes required by the cost, we introduce a real scaling parameter $\theta_0$:

\begin{equation}
    \ket{f_{\boldsymbol{\theta}}} = \theta_0 U(\boldsymbol{\theta}) \ket{\mathbf{0}}\,\,.
\end{equation}
Thus, despite the ket notation, $\ket{f_{\boldsymbol{\theta}}}$ should be understood as a scaled embedding of the discretized field rather than a normalized quantum state. The scalar $\theta_0$ can be optimized alongside $\boldsymbol{\theta}$ (and, in practice, absorbed into measurement rescaling within the QNPU layer) so that the residual in Eq.~\eqref{cost_burgers} is evaluated with the appropriate amplitude.

\subsubsection{QNPU - Realizing and measuring the cost function}
The QNPU is the layer of quantum gates responsible for realizing the cost function. It evaluates nonlinear terms acting on the trial function encoded in the quantum state and, as mentioned previously, the measurement of an ancilla qubit directly yields the squared norm of the residual. To construct such gates, it is useful to recast the cost function in a form more amenable to quantum evaluation in a circuit, as shown in Fig. \ref{circuit_dieter}.

Consider the state $\ket{f_{\boldsymbol{\theta}}} = \boldsymbol{\theta}_0U(\boldsymbol{\theta})\ket{\mathbf{0}}$ as the state encoding the function in the desired step $f(t+\tau)$, and $\ket{\tilde{f}} = \tilde{\boldsymbol{\theta}_0}\tilde{U}\ket{\mathbf{0}}$ the state encoding the function in the previous time step, i.e. $f(t)$, where $\tilde{U} = U(\tilde{\boldsymbol{\theta})}$ and the parameters $\tilde{\boldsymbol{\theta}}$ correspond to the configuration in the previous step. Substituting these into Eq. \eqref{cost_burgers} we get:
\begin{align}
\label{cost_burgers_qnpu}
\lVert \ket{f_{\boldsymbol{\theta}}} - (I + \tau O)\ket{\tilde{f}} \rVert _{2}^2
&= \left(\bra{f_{\boldsymbol{\theta}}} - \bra{\tilde{f}}(I+\tau O)^{\dagger} )
   (\ket{f_{\boldsymbol{\theta}}} - (I+\tau O)\ket{\tilde{f}}\right) \notag \\
&= \lvert \theta_{0} \rvert ^2 
   - \theta_{0}^*\tilde{\theta}_{0}\bra{\mathbf{0}}U^\dagger(\boldsymbol{\theta}) (I-\tau O)\tilde{U}\ket{\mathbf{0}} \\ &\quad
   - \tilde{\theta}_{0}^*\theta_{0}\bra{\mathbf{0}}\tilde{U}^\dagger (I+\tau O)U(\boldsymbol{\theta})\ket{\mathbf{0}} \notag \\
&\quad + \lvert \tilde{\theta}_{0} \rvert ^2 
   \bra{\mathbf{0}}\tilde{U}^\dagger (1+\tau O)^{\dagger}(I+\tau O)\tilde{U}\ket{\mathbf{0}} \notag \\
&= \lvert \theta_{0} \rvert ^2
   - 2 \operatorname{Re}\!\left( \tilde{\theta}_{0}\theta_{0}\bra{\mathbf{0}} \tilde{U}^{\dagger}(I + \tau O)U(\boldsymbol{\theta})\ket{\mathbf{0}} \right) + K\,\,,
\end{align}
where the final term $K$ is constant with respect to $\boldsymbol{\theta}$ and therefore irrelevant for gradient-based optimization.

\begin{figure}
    \centering    \includegraphics[width=0.8\linewidth]{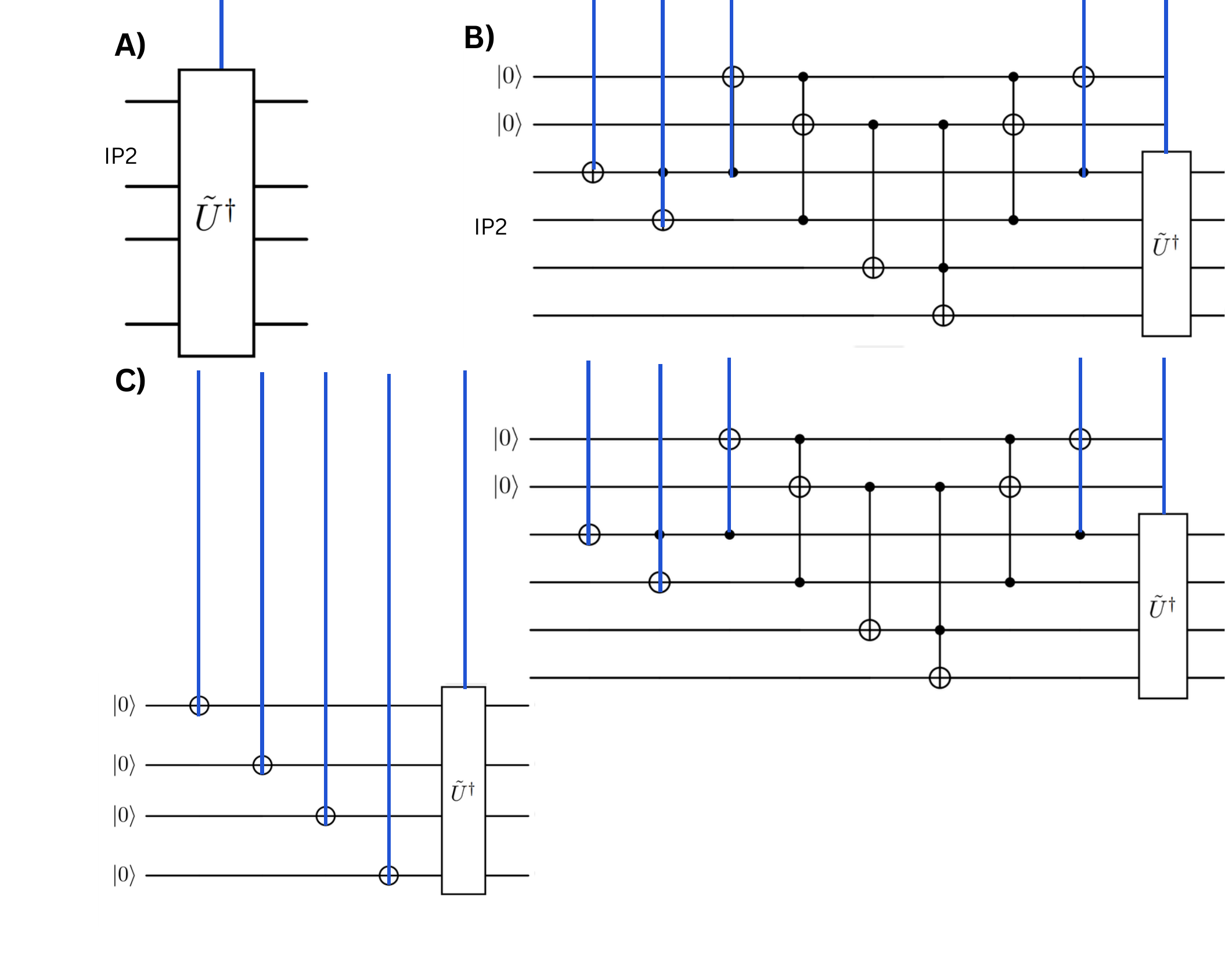}
    \caption{Example of QNPU for Burgers equation, realizing the terms pertinent to equation \eqref{cost_burgers_qnpu}. All blue lines indicate that the gate is controlled by the ancilla qubit, on which, upon measurement, approximates \eqref{cost_burgers_qnpu}. \textbf{A)} Responsible for term $\bra{\mathbf{0}} \tilde{U}^{\dagger}U(\boldsymbol{\theta})\ket{\mathbf{0}}$. \textbf{B)} A circuit $\mathcal{A}$, that shifts bits of the state (as an adder circuit~\cite{vlatko_quantum_arithmetic_1996}), rendering it as an expectation value $\bra{\mathbf{0}}\tilde{U}^\dagger \mathcal{A}U(\boldsymbol{\theta})\ket{\mathbf{0}}$. \textbf{C)} The adder circuit can then be used to calculate the derivative.}
    \label{burgers_qnpu}
\end{figure}

The QNPU must capture the action of linear operators $O_i$, which in general are not unitary. A standard approach is to decompose $O_i$ as a linear combination of Pauli strings~\cite{lubasch_variational_2020}:

\begin{equation}
O_{i} = \sum_{j=1}^m c_{j}\, \mathbf{p}_{j}, 
\quad c_{j} \in \mathbb{C}, 
\quad \mathbf{p}_{j} \in \{ \sigma_{j_1}\otimes \cdots \otimes \sigma_{j_n} \mid j_k \in \{0,1,2,3\} \},
\label{op_dec}
\end{equation}
where $\sigma_{0}=I$, $\sigma_{1}=X$, $\sigma_{2}=Y$, $\sigma_{3}=Z$. Since the Pauli strings form a basis for operators on $(\mathbb{C}^2)^{\otimes n}$, any $O_i$ can be expressed in this way. In practice, however, the number of terms $m$ must be kept small to avoid deep and costly circuits. Figure~\ref{burgers_qnpu} illustrates a QNPU designed to implement the Burgers cost function.

Nonlinear terms such as powers of the trial function can be implemented using CNOT-based constructions that realize pointwise products, thereby introducing additional nonlinearity into the output. Finite-difference derivative operators may be implemented via shifts that, once fed into the classical optimizer, realize finite-difference stencils, as detailed in the supplementary material of~\cite{lubasch_variational_2020}.

The read-out of the QNPU is performed using the ancilla qubit (see Fig.~\ref{circuit_dieter}), following the Hadamard test~\cite{luongo_chapter_nodate}. This procedure ensures that the expectation value of the ancilla qubit in the computational basis behaves as a random variable that serves as an estimator for functions such as $\mathcal{C}$. Then we get the sum $\sum_{i}\text{Re}(\mathcal{C}(x_{i}))$, i.e., the sum of all possible real values of the cost function.

The QNPU framework illustrates how quantum algorithms can leverage memory efficiency to encode high-dimensional vectors and exploit entanglement to capture correlations, such as cross-scale features in CFD, that are not quantum in origin. Ancilla-assisted measurements mitigate read-out bottlenecks by converting nonlinear cost evaluations into efficient expectation-value estimates. Although this specific setup has yet to be fully applied to CFD, proof-of-principle demonstrations (e.g., solving a 1D time-independent nonlinear Schrödinger equation on the IBM Quantum platform~\cite{lubasch_variational_2020}) confirm the viability of the approach.

\section{Physics-Informed Neural Networks and Quantum Physics-Informed Neural Networks}
\label{QNNs_and_QPINN}
\subsection{PINNs} \label{pinn}
Physics-Informed Neural Networks (PINNs) are a model of supervised learning where a given neural network is trained to approximate the solution of a PDE while being required to satisfy physical laws~\cite{raissi_physics-informed_2019}. In order to better understand the model, let $u: \Omega \times [0,T] \to \mathbb{R}^m$ be an at least differentiable function, where $\Omega \subset \mathbb{R}^d$ corresponds to its spatial domain, and $[0, T]\in \mathbb{R}$ is a time interval, that for a given nonlinear differential operator $\mathcal{P}(\cdot)$ satisfies 
\begin{equation}
\label{PINN_problem}
\mathcal{P}(u(x,t)) = 0, \quad x \in \Omega ,  t\in[0, T]\,\,.
\end{equation}

Given that the solution of a PDE must be at least differentiable, supported by the neural network approximation theorem~\cite{hornik_multilayer_1989}, we can build an approximation $u_{\boldsymbol{\theta}}$ where $\boldsymbol{\theta} \in \mathbb{R}^K$ is a parameter vector of weights and biases of the neural network. Typically, for this problem, a set of data points $\{ x^{(i)}, t^{(i)}\}_{i=1}^N$ is required, where the evaluation of the function $u(x^{(i)}, t^{(i)})$ is known. The approximation problem consists of applying an optimization method, generally gradient descent, to minimize the mean squared error (MSE):
\begin{equation}
\label{MSE}    
\min_{\boldsymbol{\theta} \in \mathbb{R}^m} \frac{1}{N}\sum_{i=1}^N \lvert \lvert u(x^{(i)}, t^{(i)}) - u_{\boldsymbol{\theta}}(x^{(i)} ,t^{(i)})\rvert  \rvert ^2\,\,.
\end{equation}
This approach has the issue of heavily relying on available data points. The PINN model circumvents this issue, leveraging the efficacy of the previously mentioned autodiff to change this optimization problem. By defining the neural network $f:= \mathcal{P}(u_{\boldsymbol{\theta}})$, the problem is one of optimization, requiring the residual of the PDE to be zero and that $f$ respects the boundary and initial conditions that properly define the uniqueness of the PDE solution. For residual evaluation, $N_{f}$ points are sampled from the domain, known as collocation points, from a given probability distribution, without the need to worry about the geometry of the mesh. The MSE loss is defined as:
\begin{equation}
\label{residue}    
\mathcal{L}_{f}(x^{(i)}, t^{(i)}, \boldsymbol{\theta}) = \sum_{i=1}^{N_{f}} \frac{1}{N_{f}} \lvert \lvert f(x^{(i)}, t^{(i)}) \rvert  \rvert ^2\,\,.
\end{equation}
Then, considering the loss with respect to the boundary and initial conditions on the $N_{b}$ and $N_0$ points, respectively:

\begin{equation}
\label{bc_loss}
\mathcal{L}_{\text{BC}}(x^{(i)}, t^{(i)}, \boldsymbol{\theta}) = \sum_{i=1}^{N_{b}} \frac{1}{N_{\text{b}}} \lvert \lvert u(x^{(i)}, t^{(i)}) - u_{\boldsymbol{\theta}}(x^{(i)}, t^{(i)}) \rvert  \rvert ^2\,\,,
\end{equation}

\begin{equation}
\label{initial_condition_loss}    
\mathcal{L}_{0}(x^{(i)}, t^{(i)}, \boldsymbol{\theta}) = \sum_{i=1}^{N_{0}} \frac{1}{N_{0}} \lvert \lvert u(x^{(i)}, 0) - u_{\boldsymbol{\theta}}(x^{(i)}, 0) \rvert  \rvert ^2\,\,,
\end{equation}
the minimization becomes:
\begin{equation}
\label{total_loss} 
\mathcal{L}(x^{(i)}, t^{(i)}, \boldsymbol{\theta}) = \min_{\boldsymbol{\theta} \in \mathbb{R}^m} (\lambda_{1}\mathcal{L}_{f} +\lambda_{2} \mathcal{L}_{0} + \lambda_{3} \mathcal{L}_{\text{BC}}) \,\,,
\end{equation}
where $\lambda_{1}, \lambda_{2}, \lambda_{3}$ are weights for each loss term. During training, the weights cause the model to prioritize certain features over others. A larger weight for $\mathcal{L}_{\text{BC}}$ favors that boundary conditions are respected.

PINNs are just establishing themselves as a new cutting-edge methodology. Although many of their properties and guarantees of stability and convergence are still being studied~\cite{gazoulis_stability_2025}, PINNs show promising results in quantitative inferences for fluid dynamics in very geometrically intricate domains~\cite{raissi_hidden_2018} and in their ability to predict physical quantities of interest in turbulent flows~\cite{eivazi_physics-informed_2022}.

Along with integration of complex physics, such as dealing with stochastic phenomena and incorporating complex boundary conditions in the model, a big challenge for PINNs is still high dimensionality. Precise models may require a very large number of parameters, increasing computational complexity. This is where incorporating quantum routines becomes a favorable course of action. By leveraging the expressivity of quantum circuits, it is possible to create models with one tenth of the number of parameters with similar accuracy~\cite{farea_qcpinn_2025}, as discussed in the next section.

\subsection{Quantum PINN}

Quantum machine learning is a broad area of study, consisting of approaches with different architectures for different models of quantum computation. Among these, the most prominent work on adapting physics-informed learning is done via circuit-based models. The procedure consists in optimization of parametrized quantum circuits via VQAs (see section \ref{vqa}) that include data of the task in the loss function. In addition, by incorporating nonlinearity in a layered structure via quantum feature maps, a machine learning model \emph{analogous} to neural networks is obtained, but one that runs in a hybrid quantum-classical algorithm - Quantum Neural Networks (QNNs).

It is important to stress that QNNs are \emph{analogous} to neural networks. While the name may suggest otherwise, QNNs are not merely quantum counterparts; rather, they constitute machine learning models in their own right. QNNs define their own family of functions and have their own universal approximation properties~\cite{goto_universal_2021, mangini_variational_2023, Jaderberg2024}. Nonetheless, they share some similarities with classical neural networks, particularly their ``layered structure'' and combination of linear and nonlinear operations. Although the literature can fall short of providing a precise definition for Quantum Physics-Informed Neural Networks (QPINNs), we now look at structures common to most current QPINN models and attempt to properly define their framework.

\subsubsection{Explicit linear quantum models}
Explicit linear quantum models play for QNNs a role similar to that of weight matrices in classical neural networks.

In supervised machine learning tasks, we typically work with a dataset $Z = \{ x^{(i)}, y^{(i)} \}_{i=1}^M$ where each $x^{(i)} \in \mathcal{X}$ is a feature vector and $y^{(i)} \in Y$ is the corresponding target value, given by some underlying function $f: X \to Y$:
\begin{equation}
y^{(i)} = f(x^{(i)})\,.
\end{equation}
For example, in the case of PINNs (see section \ref{pinn}), the input vectors $x^{(i)}$ are coordinates in the solution domain, while the outputs $y^{(i)}$ may correspond either to evaluations of the solution at boundary conditions or to residuals at collocation points.

 For a dataset $Z$, let $U(x, \boldsymbol{\theta})$ be a quantum circuit that depends on the input features $x \in X$ and a parameter vector $\boldsymbol{\theta} \in \mathbb{R}^K$. Acting on the all-zeros state of the computational basis, it produces the feature-dependent state $\ket{\psi(x, \boldsymbol{\theta})} = U(x, \boldsymbol{\theta})\ket{\mathbf{0}}$. Let $A$ be a self-adjoint operator representing an observable. The linear quantum model is then defined as the expectation value:
\begin{equation}
\label{linear_quantum_model}    
f_{\boldsymbol{\theta}}(x) = \bra{\psi(x, \boldsymbol{\theta})} A \ket{\psi(x, \boldsymbol{\theta})}\,\,, 
\end{equation}
or, equivalently, in terms of the density operator $\rho = \ket{\psi(x, \boldsymbol{\theta})}\bra{\psi(x \boldsymbol{\theta})}$:
\begin{equation}
\label{density_model}
f_{\boldsymbol{\theta}} = \mathrm{Tr}(A\rho)\,\,.
\end{equation}
The model is trained as a VQA so that for every sample $x^{(i)} \in Z$, we have $f_{\boldsymbol{\theta}}(x^{(i)}) \approx f(x^{(i)})$. The objective is to obtain a good approximation 
$f_{\boldsymbol{\theta}}(x) \approx f(x)$ 
even for inputs $x$ outside the training dataset, thereby ensuring generalization.

The analogy with linear models in neural networks becomes clearer when noting that the linear quantum model can be expressed as a Hilbert–Schmidt inner product between the state $\rho$ and the observable $A$: $\langle A, \rho \rangle_{\text{HS}}:= \mathrm{Tr}(A \rho)$, which parallels the role of the perceptron in classical settings.

It is often convenient to decompose the circuit $U(x, \boldsymbol{\theta})$ as a separable operator in two blocks: a unitary $W(x)$ which embeds the features into quantum states and a variational block $V(\boldsymbol{\theta})$ consisting of gates parameterized by the trainable parameters.

\subsubsection{Quantum Neural Networks and nonlinearity}

The class of functions representable by the linear model is too restrictive. After all, a quantum circuit is only universal in the sense it can implement arbitrary unitary evolutions, but not universal function approximations~\cite{schuld_effect_2021}.

Drawing inspiration from universal approximation properties in classical machine learning, a natural way to enhance expressivity is by introducing nonlinearity into the model. While quantum circuits are defined by linear unitary maps, non-linearity already arises when translating classical data into quantum states. For example, amplitude encoding (see section \ref{encoding_on_vqa}) provides a nonlinear mapping from feature vectors $x$ to quantum states. More generally, quantum feature maps apply a nonlinear transformation $\Phi$ to the data and then embed it into a high-dimensional Hilbert space as a quantum state via $W_{\Phi(x)}\ket{\mathbf{0}}$. Such feature maps play a key role in establishing universal approximation properties for quantum machine learning models~\cite{goto_universal_2021, perez-salinas_data_2020}.

There exist many procedures for encoding classical data into quantum states, each with distinct advantages and drawbacks~\cite{rath_quantum_2024, ranga_quantum_2024}. A particularly relevant class of data-encoding gates with well-characterized approximation properties is that of the form
\begin{equation}
\label{data_encoding}
    W(x_j) = e^{-ix_jG}, \quad G = G^\dagger\,\,.
\end{equation}

As shown in~\cite{schuld_evaluating_2019, mangini_variational_2023}, expectation values of the form in Eq. \eqref{linear_quantum_model} with such gates can be expressed as partial Fourier series:
\begin{equation}
\label{fourier_truncated}
   f_{\boldsymbol{\boldsymbol{\theta}}}(\boldsymbol{x})  = \sum_{\boldsymbol{\omega} \in \Omega} c_{\boldsymbol{\omega}} e ^{-i \boldsymbol{\omega} \cdot \boldsymbol{x}}\,\,,
\end{equation}
where $\Omega \subset \mathbb{R}^L$ is a frequency spectrum determined by the number $L$ of data-encoding operations in the circuit, and the coefficients $c_{\boldsymbol{\omega}}$ depend on the variational gates and the observable. By reuploading data, the set $\Omega$ expands, yielding richer Fourier representations. Since Fourier series can approximate a wide class of functions, including all square-integrable ones~\cite{al-gwaiz_fourier_2008}, this motivates the ``formal'' definition of a QNN~\cite{mangini_quantum_2021} as a parametrized quantum circuit with layered data reupload structure:
\begin{equation}
    \label{data_reupload}
   \prod_{l=1} ^L V_l(\boldsymbol{\boldsymbol{\theta}}_l)W_l(x)\,\,.
\end{equation}

Among encoding strategies, the most common in QPINNs is angle encoding, which takes the components of a feature vector $x = (x_j)_{j=1}^m$ as arguments of rotation gates:
\begin{equation}
\label{angle}
   \ket{\psi(x)}  = \bigotimes_{j=1}^mR_{\sigma}(x_{j})\ket{\mathbf{0}}  = \bigotimes_{j=1}^m e^{-i \frac{x_{j}}{2} \sigma }\ket{\mathbf{0}}\,\,,  
\end{equation}
where $\sigma \in \{X, Y, Z \}$ is a Pauli-gate.

Although this review focuses on qubit-based quantum computing, it is worth noting that QNNs can also be implemented in continuous-variable (CV) quantum computing~\cite{killoran_continuous-variable_2019, bangar_experimentally_2023}. QPINNs have been explored in CV models~\cite{markidis_physics-informed_2022, panichi_quantum_2025}, though~\cite{farea_qcpinn_2025} reports greater training instability compared to qubit-based approaches.

\subsubsection{Quantum PINNs via QNN}

One of the earliest works in this area is the Differentiable Quantum Circuit (DQC) proposed in~\cite{kyriienko_solving_2021}. Here, a quantum circuit represents the solution of an ordinary differential equation (ODE), or a system of ODEs, and its derivatives are computed via parameter-shift rules (see section \ref{gradients}) for residual evaluation.

Given a nonlinear function $\Phi: \mathbb{R}\to \mathbb{R}$, one can define a differentiable quantum feature map, such as the product feature map:
\begin{equation}
    U_{\Phi}(x) = \bigotimes_{j=1}^{m}R_{\sigma,j}(\Phi(x))\,\,,
\end{equation}
where $\sigma$ is a Pauli gate and $j$ indexes the qubits. Specific choices of $\Phi$ and $\sigma$ yield different basis functions. For example, setting $\sigma = Y$ and $\Phi(x) = 2n \arccos{x}$ with $n \in \mathbb{Z}$ produces the Chebyshev feature map. 

Once the feature map is defined, standard VQA procedures (section \ref{vqa}) are applied: an ansatz for the variational block $V(\boldsymbol{\theta})$ is chosen, and an observable is selected to define the cost function. The loss in Eq. \eqref{total_loss} is computed as the MSE at boundary points, ensuring that the residual—evaluated via parameter-shift rules—vanishes. The DQC model demonstrated good numerical precision for ODEs with complex dynamics. However, it is limited to ODEs, and higher-order derivatives require exponentially many shifted evaluations ($2^n$ for the $n$-th derivative). Still, the method is not confined to simulation and offers insights into implementing QPINNs on actual quantum hardware.

In fluid dynamics, the DQC model showed promise in the benchmark problem of flow through a convergent–divergent nozzle (as also studied in~\cite{gaitan_finding_2020}). Using a quasi-1D approximation of the inviscid Navier–Stokes equations with steady-flow assumptions, the continuity, energy, and momentum equations reduce to
\begin{align}
\label{ns_quasi_1d_1}
 \frac{d \rho}{d x}&=\frac{\rho V^2 d_x(\ln A)}{T-V^2}\,\,, \\
\label{ns_quasi_1d_2}
 \frac{d T}{d x}&=\frac{T V^2(\gamma-1) d_x(\ln A)}{T-V^2}\,\,, \\
\label{ns_quasi_1d_3}
 \frac{d V}{d x}&=-\frac{T V d_x(\ln A)}{T-V^2}\,\,,
\end{align}
where $\rho$ is the density of the gas, $T$ is its temperature, $V$ is the velocity, $A$ is the variable area of the nozzle, and $\gamma$ is the ratio of specific heats, with $x \in [0, 1)$.

Despite the stiffness of this system and its sensitivity to initial velocity, the DQC approach achieved remarkably accurate solutions using six qubits, a Chebyshev feature map, and a hardware-efficient ansatz (HEA) for the variational block. Training was split into two stages: near the nozzle throat, where divergence is problematic, the model separately captured the subsonic and supersonic regimes of the steady flow.

The work of~\cite{trahan_quantum_2024} applied QNNs to problems such as the forced spring–mass ODE and two-dimensional Poisson equations. Using Pennylane’s simulation environment, the authors employed variational layers composed of single-qubit rotation gates with trainable parameters and CNOT gates. The CNOT gates are implemented in such a pattern to generate strongly entangled states, i.e, states that all bipartitions are highly entangled so that for a multipartite system, tracing out a qubit still leads to an entangled state. This circuit architecture was first proposed in~\cite{schuld_circuit-centric_2020}, where it is argued that strongly entangling circuits capture higher correlations in data and cover more regions of the quantum system's Hilbert space. Data reuploading was implemented differently: after passing through one QNN ``node'' (a variational layer), measurement outcomes of all qubits were fed as inputs into the next QNN via angle encoding. Tests with both noiseless and noisy channels confirmed good accuracy for QPINNs using Adam’s optimizer, with fewer parameters and lower RMSE than classical PINNs. The best results used only four qubits (two nodes of two qubits each). For more complex CFD applications, however, the authors transitioned to a \emph{Hybrid} Quantum PINN (HQPINN), discussed in the next section.

\subsubsection{Hybrid Quantum PINNs}

HQPINNs represent the most common line of work for implementing PINNs with quantum algorithms~\cite{trahan_quantum_2024, sedykh_hybrid_2024, leong_hybrid_2025, farea_qcpinn_2025}. In these approaches, the quantum model is typically inserted as a layer within a multilayer perceptron (MLP), with data being pre- and postprocessed by the surrounding classical network. This setup leverages the nonlinearity of the neural network as part of the feature map, while the parametrized quantum circuit acts as a ``shortcut'' that reduces the overall number of trainable parameters. It should be noted that although these models are termed \emph{hybrid}, even ``pure'' QPINNs are hybrid in a sense, since they rely on VQAs for optimization. The term hybrid here refers specifically to the \emph{network structure}. The subsequent discussion reviews how HQPINNs have been applied.

In the HQPINN of~\cite{trahan_quantum_2024}, the network inserts quantum nodes that receive input from hidden layers of a PINN. After measuring the qubits individually, each expectation value is passed to neurons in a dense classical layer. The model was applied to Burgers’ equation (Eq. \cref{burgers_eq}) with viscosity $\nu = \tfrac{0.01}{\pi}$, Dirichlet boundary conditions, and an initial shock $u(x,0) = -\sin(\pi x)$. Using the same circuit architecture as in the purely quantum case, different configurations of nodes, variational layers, and qubits were tested. The best convergence was achieved with 5 qubits and 5 variational layers. Functionally, the QNN acted as a fifth hidden layer in a fully connected MLP with 20 neurons, feeding into a sixth hidden layer and then an output layer. With small error relative to the exact solution, the HQPINN reduced the number of parameters by more than 50\%: a comparable classical PINN required 3441 parameters, while the HQPINN achieved similar performance with only 1456. The classical components were implemented in TensorFlow with the Keras API.

In~\cite{sedykh_hybrid_2024}, a different HQPINN architecture was proposed. Here, parallel classical–quantum networks were used to solve the steady flow of an incompressible 3D fluid flow in a Y-shaped mixer, governed by the Navier–Stokes equations. Boundary conditions included no-slip walls, fixed inlet velocity, and fixed outlet pressure. As before, the QNN was implemented as a neural network layer, but some of the connected neurons that map into the QNN also connect to another neuron that runs in parallel to the quantum circuit. The outputs of these neurons, together with the qubit expectation values, were used as outputs to predict the velocity components $(v_x, v_y, v_z)$ and pressure $p$. This parallel structure was designed to improve information processing and capture richer patterns. The quantum layers consisted of single-qubit rotation gates in the variational block, entangled by CNOTs. An angle encoding procedure translated the classical network preprocessed values into the three-qubit quantum circuit. Compared to a classical PINN with identical hyperparameters, the HQPINN converged faster, achieving 21\% lower loss (and thus higher accuracy). The downsides were increased runtime, despite the small number of qubits, and some asymmetry observed between flow profiles in the mixer’s inlets.

In~\cite{leong_hybrid_2025}, another HQPINN with parallel structure was implemented. Since QNNs represent functions as truncated Fourier series (Eq. \eqref{fourier_truncated}), they may struggle with strongly non-harmonic solutions. To address this, a parallel MLP was used to complement the QNN. Domain points were fed simultaneously to both networks, with outputs combined only at the final layer: the $i$-th output was the sum of the $i$-th qubit register’s $Z$ expectation value and the $i$-th neuron’s output. The QNN employed angle encoding and a data-reuploading structure (Eq. \eqref{data_reupload}). The variational block followed the design of~\cite{trahan_quantum_2024}, using four qubits with rotational gates and CNOTs for strong entanglement.  

This HQPINN was applied to one- and two-dimensional Euler equations for steady flow:
\begin{equation}
    \label{euler_eq}
    \partial_t U+\nabla \cdot f(U)=0, x \in \Omega \subset \mathbb{R}^d, d=1,2, t \in(0, T]\,\,,
\end{equation}
with $U$ and $f(U)$ defined according to the dimension $d$ (see~\cite{mao_physics-informed_2020}). In 1D, both harmonic and contact discontinuity regimes were studied; in 2D, the case was transonic flow past a NACA0012 aerofoil, requiring adaptive gradient weighting due to its complexity. As expected, QPINNs outperformed in the harmonic case but underperformed for more challenging problems. HQPINNs matched classical models in 1D with nearly half the number of parameters, but in 2D the classical PINN remained superior, especially when accounting for quantum emulation overhead. Nonetheless, this work highlights the potential of HQPINNs for tackling complex CFD problems.

Finally,~\cite{farea_qcpinn_2025} (where HQPINNs are termed QCPINNs) provided an extensive benchmarking study across various PDEs. In this architecture, inputs were first processed by an MLP, then passed to a QNN, and finally decoded by another MLP. Both discrete- and continuous-variable quantum models were tested, with discrete-variable QNNs consistently outperforming CV models. The discrete QNN used five qubits with a hardware-efficient ansatz (HEA) based on single-qubit rotations. Four HEA variants were evaluated: alternated, cascade, layered, and cross-mesh. Each variant differs in depth, connectivity, and parameter count. The benchmarking was thorough: every PDE was tested under all HEA designs, both angle and amplitude encoding, and compared against two classical PINNs. Metrics included parameter count, solution error, and final training loss. When it comes to fluid dynamics applications, these HQPINNs were applied to the time-dependent 2D lid-cavity flow governed by the incompressible Navier-Stokes equations. Boundary conditions and other details are provided in the Appendix of~\cite{farea_qcpinn_2025}. The cascade HEA with angle encoding achieved the best performance, yielding nearly 5\% lower training loss than the densest classical model, while using only about 10\% of its parameters. However, challenges remain, such as difficulty in capturing detailed flow near the moving lid and limited improvement over classical PINNs in certain cases.

\begin{figure}
    \centering
    \includegraphics[width=1.0\linewidth]{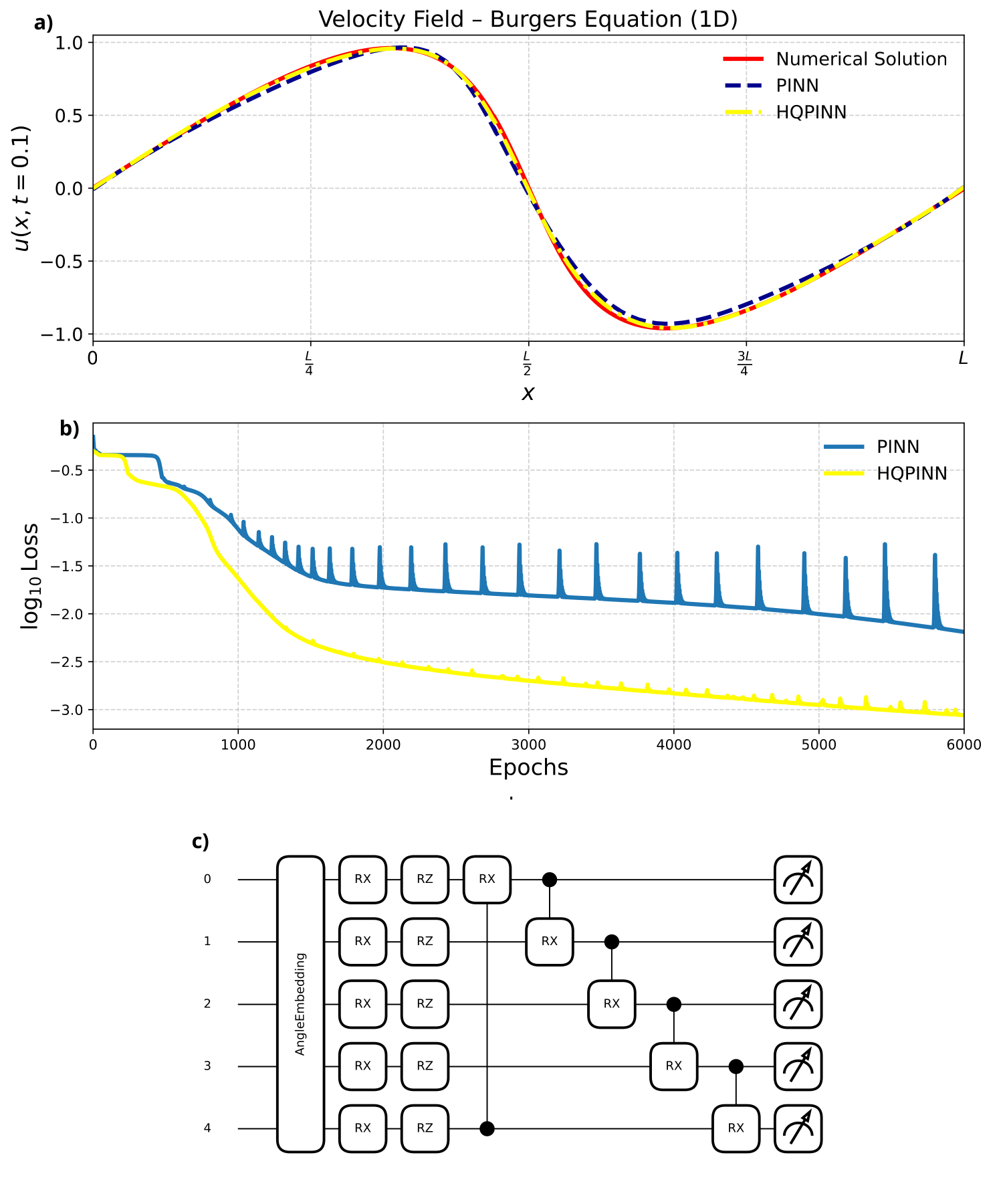}
    \caption{Example of HQPINN architecture applied to the Burgers equation with Dirichlet boundary conditions and initial condition \(u(x,0) = \sin\!\left(\tfrac{2 \pi}{L}x\right)\). The model integrates classical layers (20 neurons each) with a five-qubit angle-cascading quantum circuit~\cite{sim_expressibility_2019}. \textbf{a)} Comparison of solutions learned by HQPINN  (yellow) with numerical solution via FDM (red) and a classical PINN (blue). \textbf{b)} Comparison of loss by epochs between HQPINN (yellow) and classical PINN (blue). Compared to its classical counterpart (1341 parameters), the HQPINN (321 parameters) converges faster and attains over twice the precision within the same number of training epochs. \textbf{c)} Variational circuit employed in the quantum layer. The ansatz was not only chosen due to previous success in higher-dimensional problems~\cite{farea_qcpinn_2025} but also due to empirical testing.
}
    \label{hqpinn_model}
\end{figure}

In Figure~\ref{hqpinn_model}, we present an example of an HQPINN architecture inspired by the models discussed above. To illustrate its performance, we solve the Burgers equation~\eqref{burgers_eq} with Dirichlet boundary conditions and the initial condition \(u(x,0) = \sin\!\left(\tfrac{2 \pi}{L}x\right)\), using the space-time coordinates \((x,t)\) as input. Similar to the approach in~\cite{farea_qcpinn_2025}, the HQPINN incorporates a quantum layer, which receives inputs from two preceding classical pre-processing layers and outputs the measurement results to be post-processed by the other two classical layers. Each classical layer consists of 20 neurons. The quantum component employs the angle-cascading variational circuit~\cite{sim_expressibility_2019} with five qubits, as depicted in panel (c) of Figure~\ref{hqpinn_model}.  

For comparison, we also consider a classical PINN with the same number of layers (five), obtained by replacing the quantum layer with a dense layer of 20 neurons, as well as the finite difference method (FDM) solution. The HQPINN demonstrates a great reduction in the number of parameters, totaling 321 compared to 1341 in its classical counterpart. As shown in Figure~\ref{hqpinn_model}, despite having fewer parameters, HQPINN achieves noticeably higher accuracy, converging faster and achieving a substantially lower loss, achieving around a 2.6-times improvement in precision within the same number of training epochs

Despite these encouraging results, hybrid quantum-classical models face practical limitations that must be considered in future developments. Current implementations require a nontrivial number of qubits even for small-scale fluid dynamics problems, and the circuit depth rapidly grows with problem dimensionality, which increases exposure to quantum noise and decoherence. While hybrid training partially mitigates these effects, since the classical optimizer can adapt to hardware noise during training, it does not fully eliminate them. In practice, the optimization process may implicitly learn and compensate for systematic noise patterns, which improves convergence but also limits generalization across devices. Moreover, scaling these models to higher-dimensional or time-dependent flows remains challenging due to barren plateaus and measurement overhead. These factors emphasize the need for compact ansatze and noise-aware training strategies tailored to near-term hardware.

\section{Quantum Inspired Algorithms}
\label{quantum_insp_cesar}

CFD faces a severe scalability challenge often compared to the \textit{curse of dimensionality}, which becomes particularly evident in turbulent and high-resolution simulations. The difficulty arises from the rapid growth in the number of degrees of freedom required to discretize the spatio-temporal domain and to resolve the multiple physical variables inherent in CFD models. For instance, the number of grid-points required for the direct numerical simulation of a three-dimensional turbulent flow scales as $Re^{9/4}$, where $Re$ is the Reynolds number of the flow. The flow over a wing of a commercial airliner has $Re\sim  10^7$. Typical time-resolution requirements scale the total computational cost up to $Re^3$ (grid points x time steps)~\citep{Pope2000}. Therefore, when turbulence, multiphysics coupling, or parametric uncertainty are taken into account, the effective dimensionality of the problem increases even further, leading to computational demands that escalate dramatically. 

A similar phenomenon occurs in the study of Quantum Many-Body Systems (QMBS). Representing a full quantum state requires specifying all coefficients in the Hilbert-space basis, whose size grows exponentially with the number of particles. For example, a system of $N$ two-level particles has a Hilbert space of dimension $2^N$.

Both CFD and QMBS therefore, face exponential or near-exponential growth in computational complexity, and both fields have sought strategies to overcome these obstacles. In both cases, the underlying objects of study, fluid fields or quantum states, can be represented as high-dimensional tensors, making tensor-based compression techniques a natural tool. One promising approach is the use of Tensor Networks (TNs), which have been remarkably successful in compressing high-dimensional quantum states. More recently, TN-based methods have also been adapted to other domains, including CFD~\cite{Pisoni2025, Danis2025, Kornev2023, Holscher2025, Peddinti2024, Kiffer2023, Nikita2025, gourianov_quantum_2022, gourianov2022a} and Machine Learning~\cite{Liu2023, Wang2025, Stoudenmire2017, Newman2004, Rieser2023}.

Tensor Network methods have proven to be powerful computational tools, particularly for QMBS that obey an area law for entanglement, where correlations are mostly localized~\cite{Schollwock2011}. The central idea is to represent a high-dimensional quantum state, which would formally require exponentially many coefficients, through a decomposition into a network of lower-order tensors. This allows efficient compression of the state, thereby mitigating the curse of dimensionality. It is important to emphasize that the TN technique does not necessarily involve quantum computing, although it is possible to load such networks into quantum computers, as shown in Fig. \ref{variational_ansatz}.

Similar gains have been reported in CFD. For example, applying the Finite Element Method (FEM) within a TN formulation to solve the Navier–Stokes equations in complex geometries yields exponential reductions in memory usage and exponential speed-up compared to traditional FEM~\cite{Kornev2023}. In turbulence modeling, Nikita et al.~\cite{Nikita2025} achieved memory savings of order $10^6$ and cost reductions of order $10^3$ using a TN-based Finite Difference Method (FDM) to solve a $5{+}1$-dimensional partial differential equation (PDF) of a chemically reactive turbulent flow. Likewise, Danis et al.~\cite{Danis2024} employed a TN-based Finite Volume Method (FVM) for the Shallow Water Equations, reporting speed-ups of up to $124\times$ while maintaining the formal high-order accuracy of conventional FVM. These are only a few representative examples; additional applications are summarized in Table~\ref{tab:tn_cfd}, which also includes works focused primarily on data compression in CFD\footnote{Architectural notation follows the original articles. For further implementation details, we recommend the cited works.}.

\begin{table}[!h]
    \centering
    \renewcommand{\arraystretch}{1.15} % Adjust row spacing
    \setlength{\tabcolsep}{4pt} % Reduce column padding
    \caption{Applications of tensor networks in CFD problems. The meaning of the acronyms are: TT = Tensor Train; TT-FVM = TT Finite Volume Method; WENO-TT = Weighted Essentially Non-Oscillatory TT; TetraFEM = TT Finite Element Method; QTT = Quantized TT; TT-SFV+WENO = TT Stochastic Finite Volume with WENO; Hybrid TT-SFV = Hybrid TT Stochastic Finite Volume.}
    \label{tab:tn_cfd}
    \begin{tabular}{p{4.5cm} p{2.8cm} p{5.5cm}}
    \toprule
    \textbf{CFD Problem} & \textbf{TN Arch.} & \textbf{Reported Gains (vs classical)} \\
    \midrule
    Incompressible NS~\cite{Pisoni2025} & TT & $97.8\%$ fewer parameters; scaling polylogarithmic with resolution; energy spectrum preserved vs. DNS \\
    Shallow Water Eq.~\cite{Danis2024} & TT-FVM & Preserves $3^{\text{rd}}$–$5^{\text{th}}$ order; up to $124\times$ faster runtime \\
    Compressible Euler~\cite{Danis2025} & WENO-TT & $5^{\text{th}}$ order maintained; up to $10^3\times$ faster vs. WENO \\
    NS in T-mixer~\cite{Kornev2023} & TetraFEM & Exponential FEM matrix compression; exponential runtime speed-up \\
    2D turbulence~\cite{Holscher2025} & TT/QTT & Up to $12\times$ faster (GPU vs. CPU); bond dimension grows polylogarithmically \\
    Incompressible NS~\cite{Peddinti2024} & MPS & Memory/runtime scale logarithmically with mesh; avoids full state reconstruction \\
    Lid-driven cavity~\cite{Kiffer2023} & MPS & $97\%$ compression; up to $17\times$ faster vs. DNS \\
    Curvilinear NS~\cite{Vanhulst2025} & TT/QTT & Velocity field compressed $20\times$ (error $<0.3\%$); operators up to $10^3\times$ compressed \\
    Vorticity PDFs~\cite{Nikita2025} & MPS & $10^3\times$ faster; $10^6\times$ less memory vs. dense solvers \\
    Stochastic hyperbolic~\cite{Walton2025} & TT-SFV+WENO & TT ranks scale polylogarithmically; $\mathcal{O}(10^2$–$10^3)$ cost savings \\
    Burgers/Euler + stochastic~\cite{Dubois2025} & Hybrid TT-SFV & Feasible up to $\sim 50$ parameters (classical $N^{50}$ growth); reduces exponential to polynomial scaling \\
    Channel turbulence (DNS)~\cite{vonLarcher2019} & TT/QTT & DNS data compressed up to $10^3\times$ with low error \\
    3D Chem. reacting~\cite{Sinan2024} & TT & Species tables compressed up to $75\times$ with errors $\sim$ bilinear interpolation \\
    \bottomrule
    \end{tabular}
\end{table}

The third column of Table~\ref{tab:tn_cfd} reports the gains achieved by different TN architectures. The architecture refers to the connectivity of the low-rank tensors, i.e., how tensors are arranged and contracted. While the FVM is common in CFD, most TN-based studies to date rely on FDM discretizations. This preference is due to two factors: (i) differential operators can be written directly as Matrix Product Operators (MPOs) derived from local stencils, naturally fitting the Matrix Product State (MPS) structure; and (ii) many early works consider periodic or rectangular domains, where FDM is especially convenient, while FEM and FVM are needed for more complex geometries. In what follows, we focus on the MPS\footnote{In the mathematical literature, MPS is often called Tensor Train (TT).} and MPO formalisms, as they form the foundation for most TN-based approaches. Some of these others can even be found in~\cite{tensornetwork, Schollwock2011, Ostlund1995, Verstraete2004, Vidal2006, Vidal2008}  or in the respective cited paper.

Although individual studies differ in architecture and implementation, the general methodological framework follows a common pattern. For clarity, we present a simplified outline, emphasizing FDM-based TN methods.

\emph{Discretization}: The governing equations are defined on the spatial/temporal domains with boundary conditions. Discretization yields a high-dimensional system with many degrees of freedom, as in classical CFD. 

\emph{Tensorization}: The system variables are reorganized into a multidimensional tensor\footnote{Here, we simply treat a tensor as a multidimensional array of complex numbers.}. This tensor is then decomposed into a TN, a set of smaller tensors contracted in a chosen architecture. Common structures include:  
\begin{itemize}
    \item Matrix Product State (MPS): efficient for 1D systems that obey the area-law of entanglement; forms the backbone of the Density Matrix Renormalization Group~\cite{Ostlund1995}.
    \item Projected Entangled Pair States (PEPS): generalizes MPS to 2D and higher dimensions, proving useful for representing strongly correlated systems on lattices~\cite{Verstraete2004}.
    \item Tree Tensor Networks (TTNs): arranges tensors in a hierarchical tree structure, suitable for capturing correlations with logarithmic scaling in 1D or for multiscale problems~\cite{Vidal2006}.
    \item Multiscale Entanglement Renormalization Ansatzes (MERAs): introduces disentanglement within a hierarchical network to efficiently capture critical systems exhibiting scale-invariant entanglement~\cite{Vidal2008}.
\end{itemize}

Since MPS is the most commonly used method in CFD, we will focus only on explaining this architecture, while acknowledging that other structures can also be applied for this purpose. 
An MPS is a specific type of TN that represents a high-order tensor, typically a 1D chain of physical sites or indices, as a product of a sequence of lower-order tensors, which in this architecture is given by matrices. For a $L$-dimensional tensor $\mathcal{T}_{i_1 i_2 \dots i_L}$, where $i_k$ represents the local physical index at site $k$ with dimension $d_k$, an MPS representation takes the form,
\begin{equation}
   \mathcal{T}_{i_1 i_2 \dots i_L} = \sum_{\alpha_1, \dots, \alpha_{L-1}} A^{[1]}_{i_1 \alpha_1} A^{[2]}_{\alpha_1 i_2 \alpha_2} \dots A^{[k]}_{\alpha_{k-1} i_k \alpha_k} \dots A^{[L]}_{\alpha_{L-1} i_L},
\end{equation}
where $A^{[k]}$ are called the MPS tensors or site tensors, $\alpha_k$ are the virtual indices (also known as bond indices or auxiliary indices), which are summed over. These indices connect adjacent site tensors and determine the amount of correlation or entanglement that can be captured by the MPS. The dimension of the bond indices is known as the bond dimension (or matrix dimension). It controls the expressivity and computational cost of the MPS. A larger bond dimension allows the MPS to represent more complex correlations, but also increases computational complexity.

The approach to reorganizing and decomposing the original high-dimensional tensor into an MPS largely depends on the specific problem. For instance, in certain cases, it may even be possible to obtain an exact analytical MPS representation of the underlying functions or data, thereby avoiding the need for computationally expensive numerical decompositions. This concept is supported by the fact that various classical functions, such as trigonometric, polynomial, and certain fractal functions, are known to have exact MPS representations~\cite{Khoromskij2011, Dolgov2012, Oseledets2013}. However, such analytical representations are not universally applicable. In cases where an exact analytical MPS representation is not feasible, we typically need to rely on numerical methods to approximate the function or data as a low-rank MPS. This process often involves:
\begin{itemize}
    \item Successive Singular Value Decompositions (SVDs): this method involves the sequential application of SVDs across the tensor nodes, constructing an approximate MPS by truncating small singular values. It is a widely used technique for compressing high-dimensional arrays into the MPS format~\cite{Oseledets2011}.
    \item Cross-Approximation: these methods aim to recover a low-rank MPS representation by sampling only a small number of tensor entries. They are particularly advantageous as they avoid full SVDs, rendering them more scalable for large datasets~\cite{Osedelets2010}.
    \item Variational Optimization Algorithms: these algorithms iteratively optimize the individual MPS tensors to minimize a defined cost function, often related to physical quantities like energy~\cite{Schollwock2011, Verstraete2008}.
\end{itemize}
With this mapping, the original problem is embedded within a TN representation.

\emph{Temporal Evolution:} The time evolution or the numerical solution of PDEs is carried out directly in the TN framework. In the context of CFD, this is commonly achieved through the application of operators expressed in the TN formalism, typically using an architecture closely related to MPS, known as Matrix Product Operator (MPO). In essence, while an MPS describes a state or a data vector, an MPO describes an operation or a matrix that transforms such states or data.
For an operator $\mathcal{O}$ acting on a $L$-dimensional system, its MPO representation takes the form,
\begin{equation}
    \mathcal{O}_{i_1' i_2' \dots i_L', i_1 i_2 \dots i_L} = \sum_{\beta_1, \dots, \beta_{L-1}} W^{[1]}_{i_1' i_1 \beta_1} W^{[2]}_{\beta_1 i_2' i_2 \beta_2} \dots W^{[L]}_{\beta_{L-1} i_L' i_L}\,\,,
\end{equation}
where $W^{[k]}$ are the MPO tensors at site $k$, $i_k$, and $i_k'$ are the physical indices corresponding to the input and output degrees of freedom at site $k$, and $\beta_k$ are the virtual indices. The dimension of the virtual indices controls the complexity and accuracy of the operator representation.

Numerical methods are used to perform temporal evolution in the compressed space of the TN, which can result in a lower computational cost. For example, a numerical method for solving differential equations, such as the second-order Runge–Kutta method (RK2), can be implemented by applying multiple MPO operations to the current MPS, followed by summing the resulting MPSs to efficiently advance the solution in the compressed TN format for each step.

To keep the computational cost manageable during time evolution, truncation or compression techniques are applied to limit the growth of the bond dimension in the TN. These procedures enable long-time CFD simulations while retaining the essential physical features of the solution. In practice, several compression strategies can be employed, with SVD-based truncation being the most common. After each time step, the tensor is decomposed, and small singular values are discarded, thereby reducing the bond dimension while preserving the dominant structure of the solution.

\emph{Decoding}: The final step in CFD simulations using TN is to extract classical information from the compressed representation. This typically involves reconstructing the physical fields from the MPS format by performing tensor contractions. In practice, this can be done either locally or globally. When the network has been very compressed, this may also involve interpolation or smoothing techniques to recover the original solution with good approximation. The computational cost of this decoding process is significantly lower than the full simulation, especially if only a subset of the data is required.

In order to facilitate understanding of TN applications in CFD, we present a very simple illustrative example. Here, we focus primarily on the mathematical formulations, but we also provide a code implementation for readers interested in the computational simulation. The source code is available online.

Consider the 1-D viscous Burgers equation~\eqref{burgers_eq}, where the domain is periodic, with initial condition $u(x,0) = \sin{(2\pi x)}$, defined on $x \in [0,1]$ and time interval $t \in [0,T]$.

The explicit Euler method is used for temporal evolution \cite{Burden}. With a discretized time step $\Delta t$, the update rule is,
\begin{equation}
    u_{t+1} = u_t + \Delta t\, \mathcal{L}(u_t)\,\,,
    \label{eq:euler}
\end{equation}
with the spatial operator $\mathcal{L}$ incorporating both nonlinear advection and diffusion,
\begin{equation}
    \mathcal{L}(\mathbf{u}_t) = -\mathbf{u}_t \cdot \frac{\partial \mathbf{u}_t}{\partial x} + \nu \frac{\partial^2 \mathbf{u}_t}{\partial x^2}\,\,.
    \label{eq:SP_burgers}
\end{equation}
Our goal is to solve this equation using TNs in the four steps described above. The nonlinear term introduces additional complexity, as the multiplication of tensors must be handled carefully in the MPS representation.

\paragraph{Step 1: Discretization}
\hspace{5mm}The spatial domain is discretized using $N = 2^L$ points and spacing $\Delta x = \frac{1}{N}$, yielding a one-dimensional vector $\mathbf{u}(t) \in \mathbb{R}^N$ with periodic boundary conditions.
Spatial derivatives are approximated by finite difference schemes. The first-order derivative (advection term) is approximated by the second-order centered scheme,
\begin{equation}
    \frac{\partial u_i}{\partial x} \approx \frac{u_{i+1} - u_{i-1}}{2\Delta x}\,\,,
    \label{eq:first_d}
\end{equation}
which leads to the first-derivative matrix $D^{(1)} \in \mathbb{R}^{N \times N}$ with periodic boundary conditions,
\begin{equation}
    D^{(1)} = \frac{1}{2\Delta x}
    \begin{bmatrix}
    0 & 1 & 0 & \cdots & 0 & -1 \\
    -1 & 0 & 1 & \cdots & 0 & 0 \\
    0 & -1 & 0 & \cdots & 0 & 0 \\
    \vdots & \vdots & \vdots & \ddots & \vdots & \vdots \\
    0 & 0 & 0 & \cdots & 0 & 1 \\
    1 & 0 & 0 & \cdots & -1 & 0
    \end{bmatrix}\,\,.
    \label{eq:d1}
\end{equation}
The diffusion term is approximated by the second-order centered finite difference,
\begin{equation}
    \frac{\partial^2 u_i}{\partial x^2} \approx \frac{u_{i-1} - 2u_i + u_{i+1}}{(\Delta x)^2}\,\,.
    \label{eq:second_d}
\end{equation}
Above, the subscripts refer to the discrete spatial index. This leads to the discrete Laplacian operator, represented by a circulant matrix $D ^{(2)}\in \mathbb{R}^{N \times N}$,
\begin{equation}
     D ^{(2)} = \frac{1}{(\Delta x)^2}
    \begin{bmatrix}
    -2 & 1 & 0 & \cdots & 0 & 1 \\
    1 & -2 & 1 & \cdots & 0 & 0 \\
    0 & 1 & -2 & \cdots & 0 & 0 \\
    \vdots & \vdots & \vdots & \ddots & \vdots & \vdots \\
    0 & 0 & 0 & \cdots & -2 & 1 \\
    1 & 0 & 0 & \cdots & 1 & -2
    \end{bmatrix}\,\,.
    \label{eq:D}
\end{equation}

The nonlinear advection term $u \cdot \partial_x u$ in the discretized space is evaluated through the Hadamard (element-wise) product between $\mathbf{u}$ and $D^{(1)} \mathbf{u}$,
\begin{equation}
    \left[\mathbf{u} \cdot \frac{\partial \mathbf{u}}{\partial x} \right]_i \approx u_i \cdot (D^{(1)} \mathbf{u})_i\,\,.
    \label{eq:hadamard}
\end{equation}

After discretization of the operators and the spatial variables, we now proceed to represent them in the TN formalism, using MPS for state vectors and MPOs for differential operators.

\paragraph{Step 2: Tensorization}
\hspace{5mm}The vector $\mathbf{u}$ is reshaped into a high-order tensor $\mathcal{U}_{i_1i_2\ldots i_L} \in \mathbb{R}^{2\times2\times \cdots\times 2}$ with $L$ indices, where each $i_k \in \{0,1\}$. This binary structure is introduced so that the mesh points can be mapped naturally to quantum states via binary encoding. In other words, each entry of $\mathbf{u}$, originally indexed by an integer $j \in [0, N-1]$, is associated with a unique binary string of length $L$, i.e., $j=(i_1, i_2, \ldots, i_L)_2$.  Hence,
\begin{equation}
    \mathcal{U}_{i_1 i_2 \ldots i_L} := u_j = u_{(i_1 i_2 \ldots i_L)_2}.
\end{equation}
For example, if $L=3$, we have the tensor $\mathcal{U}_{i_1 i_2 i_3} \in \mathbb{R}^{2 \times 2 \times 2}$ in terms of the vector $\mathbf{u}$,
\begin{equation*}
    \begin{aligned}
        \mathcal{U}_{000} &= u_0, \quad & \mathcal{U}_{100} &= u_4 \\
        \mathcal{U}_{001} &= u_1, \quad & \mathcal{U}_{101} &= u_5 \\
        \mathcal{U}_{010} &= u_2, \quad & \mathcal{U}_{110} &= u_6 \\
        \mathcal{U}_{011} &= u_3, \quad & \mathcal{U}_{111} &= u_7 .
    \end{aligned}
\end{equation*}

At this stage, the tensor can be decomposed by successive SVDs. First, reshape $\mathcal{U}_{i_1 i_2\ldots i_L} \in \mathbb{R}^{2 \times 2 \times\ldots \times 2}$ into a matrix $\mathcal{U}^{[1]}_{i_1, (i_2 \ldots i_L)} \in \mathbb{R}^{2 \times 2^{L-1}}$ and perform an SVD:
\begin{equation}
    \mathcal{U}^{[1]}_{i_1, (i_2 \ldots i_L)} = U^{[1]}_{i_1, \alpha_1} S^{[1]}_{\alpha_1 \alpha_1} V^{[1] T}_{\alpha_1, (i_2\ldots i_L)}\,\,.
\end{equation}
The first MPS core is then defined as,
\begin{equation}
    A^{[1]}_{i_1, \alpha_1} := U^{[1]}_{i_1, \alpha_1} \in \mathbb{R}^{2 \times r_1}\,\,,
\end{equation}
where $r_1 \leq \min(d_1, d_2 \cdots d_L)$ is the rank of the reshaped matrix, with $d_j$ the dimension of index $i_j$. In this binary case, $r_1 \leq \min(2, 2^{L-1}) = 2$.  

Next, compute $S^{[1]}  V^{[1] } \in \mathbb{R}^{r_1 \times 2^{L-1}}$ and reshape it into a new matrix $\mathcal{U}^{[2]}_{(\alpha_1 i_2), (i_3 \ldots i_L)} \in \mathbb{R}^{(r_1  \cdot 2) \times 2^{L-2}}$. Perform a second SVD:
\begin{equation}
    \mathcal{U}^{[2]}_{(\alpha_1 i_2), (i_3 \ldots i_L)} = U^{[2]}_{(\alpha_1 i_2), \alpha_2} S^{[2]}_{\alpha_2} V^{[2] T}_{\alpha_2, (i_3\ldots i_L)}\,\,,
\end{equation}
and define the second MPS by reshaping $U^{[2]}$:
\begin{equation}
    A^{[2]}_{\alpha_1, i_2, \alpha_2} := U^{[2]}_{(\alpha_1 i_2), \alpha_2} \in \mathbb{R}^{r_1 \times 2 \times r_2}\,\,,
\end{equation}
where $r_2 \leq \min(r_1 \cdot d_2, d_3\ldots d_L)$. In our case, $r_2 \leq \min(2 \cdot 2, 2^{L-2})$.

This process is repeated for each subsequent $k$-index. Specifically, reshape $S^{[k-1]} V^{[k-1]} \in \mathbb{R}^{r_{k-1} \times 2^{L - (k - 1)}}$ into matrix
$\mathcal{U}^{[k]}_{(\alpha_{k-1} i_k), (i_{k+1} \ldots i_L)} \in \mathbb{R}^{r_{k-1} \cdot 2 \times 2^{L-k}}$, perform the SVD,
\begin{equation}
    \mathcal{U}^{[k]} = U^{[k]}_{(\alpha_{k-1} i_k), \alpha_k} S^{[k]}_{\alpha_k} V^{[k] T}_{\alpha_k, (i_{k+1} \ldots i_L)},
\end{equation}
and define the $k$-th MPS core as,
\begin{equation}
    A^{[k]}_{\alpha_{k-1}, i_k, \alpha_k} := U^{[k]}_{(\alpha_{k-1} i_k), \alpha_k} \in \mathbb{R}^{r_{k-1} \times 2 \times r_k}\,\,.
\end{equation}

Finally, when $k = L$, one obtains a matrix $\mathcal{U}^{[L]} \in \mathbb{R}^{r_{L-1} \cdot 2 \times 1}$, which is reshaped into the last core,
\begin{equation}
    A^{[L]}_{\alpha_{L-1}, i_L} := \mathcal{U}^{[L]}_{(\alpha_{L-1}, i_L)} \in \mathbb{R}^{r_{L-1} \times 2 \times 1}\,\,.
\end{equation}
The full tensor can then be reconstructed as
\begin{equation}
    \mathcal{U}_{i_1 i_2 \ldots i_L} = \sum_{\alpha_1, \ldots, \alpha_{L-1}} A^{[1]}_{i_1, \alpha_1} A^{[2]}_{\alpha_1, i_2, \alpha_2} \cdots A^{[L]}_{\alpha_{L-1}, i_L}\,\,.
\end{equation}
Here, the $r_i$ are the bond dimensions that connect successive tensors. In practice, they can be truncated to $r'_i \leq \chi < r_i$, reducing memory requirements and accelerating computations at the cost of a controlled approximation.

This procedure is known as the \emph{Left-Canonical Decomposition}~\cite{Schollwock2011}. While not always the most efficient variant of SVD-based tensor decomposition, it provides a clear pedagogical example of how an MPS is constructed.

In the case of periodic boundary conditions (PBCs), where the first and last indices are connected, the open-boundary MPS decomposition described above must be modified. Instead of representing the tensor as a chain with open ends, we contract the final core with the first, forming a closed loop. This results in a periodic MPS, written as
\begin{equation}
    \mathcal{U}_{i_1 i_2 \ldots i_L} = \mathrm{Tr}\left( A^{[1]}_{i_1} A^{[2]}_{i_2} \cdots A^{[L]}_{i_L} \right)\,\,,    
\end{equation}
where the trace connects the first and last virtual indices. An equivalent strategy is to impose periodicity directly at the operator level: by defining an MPO with PBCs, the periodic structure is enforced automatically when the MPO acts on the initial MPS.

While effective, the successive SVD procedure can be computationally expensive. Several alternative techniques have been proposed~\cite{Schollwock2011}. However, as previously mentioned, one approach with virtually no computational cost is the exact MPS representation, where the mapping is performed directly. Below, we illustrate this method for derivative operators using MPOs.

Given an MPS for the initial condition, the next step is to construct the derivative operator as an MPO. Although an MPO could be derived through SVD, this is rarely efficient. Fortunately, for many cases of practical interest, such as finite-difference approximations of first and second derivatives, compact exact MPO forms exist.

For PBCs, the central finite-difference stencil for the first derivative \eqref{eq:first_d} gives the operator
\begin{equation}
    D^{(1)} = \sum_{i=1}^{L} \left( \hat{I}_1 \otimes \cdots \otimes \hat{h}^{(1)}_{i,i+1} \otimes \cdots \otimes \hat{I}_L \right)\,\,,
\end{equation}
with site $L+1 \equiv 1$ by periodicity. The two-site local operator is
\begin{equation}
    \hat{h}^{(1)}_{i,i+1} = T^- \otimes T^+ - T^+ \otimes T^-\,\,,
\end{equation}
whereas the second derivative \eqref{eq:second_d} becomes
\begin{equation}
    D^{(2)} = \sum_{i=1}^{L} \left( \hat{I}_1 \otimes \cdots \otimes \hat{h}^{(2)}_{i,i+1} \otimes \cdots \otimes \hat{I}_L \right)\,\,,
\end{equation}
with local interaction
\begin{equation}
    \hat{h}^{(2)}_{i,i+1} = -2 I \otimes I + T^- \otimes T^+ + T^+ \otimes T^-\,\,.
\end{equation}
Here $T^+ = \ket{0}\bra{1}$ and $T^- = \ket{1}\bra{0}$ are shift operators in the local computational basis.

Both $D^{(1)}$ and $D^{(2)}$ admit exact MPO representations with bond dimension 3. Let $W^{[k](n)} \in \mathbb{R}^{3 \times 3 \times 2 \times 2}$ be the MPO tensor at site $k$ for derivative order $n$, with physical indices $i_k, j_k \in \{0,1\}$ and bond indices $\alpha_{k-1}, \alpha_k \in \{1,2,3\}$. The full operator is
\begin{equation}
    \mathcal{D}^{(n)} = \sum_{\{i_k\}, \{j_k\}} \mathrm{Tr} \left[ W^{[1](n)}_{i_1 j_1} W^{[2](n)}_{i_2 j_2} \cdots W^{[L](n)}_{i_L j_L} \right] \ket{i_1 \cdots i_L} \bra{j_1 \cdots j_L},
\end{equation}
where the trace over bond indices enforces PBCs.

The MPO tensors for the first derivative are
\begin{align}
    W^{[1]} &= \left[ 0, \; T^-, \; I \right], \\
    W^{[L]} &= \left[ I, \; -T^+, \; 0 \right]^T, \\
    W^{[k]} &= 
    \begin{bmatrix}
        I & 0 & 0 \\
        -T^+ & 0 & 0 \\
        0 & T^- & I
    \end{bmatrix}, \quad 2 \leq k \leq L-1\,\,,
\end{align}
and for the second derivative:
\begin{align}
    W^{[1]} &= \left[ -2I, \; T^-, \; I \right], \\
    W^{[L]} &= \left[ I, \; T^+, \; -2I \right]^T, \\
    W^{[k]} &= 
    \begin{bmatrix}
        I & 0 & 0 \\
        T^+ & 0 & 0 \\
        -2I & T^- & I
    \end{bmatrix}, \quad 2 \leq k \leq L-1\,\,.
\end{align}

To scale these operators correctly, $\mathcal{D}^{(1)}$ is multiplied by $1/(2\Delta x)$ and $\mathcal{D}^{(2)}$ by $1/\Delta x^2$. The scaling factor may be applied globally to a single tensor or distributed locally, e.g. by multiplying each core by $\nu^{1/L}$.

The cost of constructing these MPOs is practically zero compared to the SVD-based decomposition. For further discussion on the construction and interpretation of MPOs, we refer the reader to~\cite{Kazeev2012, Garcia2021, Verstraete2008, Schollwock2011, Orus2014}, particularly regarding the role of the trace in connecting virtual indices and enforcing periodic boundary conditions through full contraction over the tensor network.

Pointwise nonlinearities, such as the Hadamard product \eqref{eq:hadamard}, can also be represented directly in the TN framework. Given two MPS $\mathcal{U}$ and $\mathcal{V}$, their element-wise product is

\begin{equation}
    \mathcal{U}_{i_1\ldots i_L} \circ \mathcal{V}{i_1\ldots i_L} = \mathcal{W}_{i_1\ldots i_L}\,\,.
\end{equation}
Operationally, this corresponds to site-wise contraction:
\begin{equation}
W^{[k]} = U^{[k]} \otimes V^{[k]}, \quad \forall k \in {1, \ldots, L}\,\,,
\end{equation}
which multiplies the bond dimensions: $r_k^{(W)} = r_k^{(U)} \cdot r_k^{(V)}$. To avoid uncontrolled growth, SVD-based truncation is typically applied immediately after.

Similarly, the sum of two MPS, $\mathcal{U} + \mathcal{V}$, is achieved by stacking their cores in block-diagonal form:
\begin{equation}
    A^{[k]}_{\text{sum}} =
    \begin{bmatrix}
    A^{[k]}_U & 0 \\
    0 & A^{[k]}_V
    \end{bmatrix}\,\,,
\end{equation}
which increases the bond dimension additively. Again, compression is applied afterwards to prevent excessive growth.

These operations are indispensable when modeling nonlinear PDEs or combining multiple terms in an update step, for instance, in the Burgers equation. State-of-the-art TN libraries provide optimized implementations of such routines, typically with configurable truncation thresholds to balance accuracy and efficiency.

\paragraph{Step 3: Temporal Evolution}
\hspace{5mm}With the discretized evolution given in Eq.~\eqref{eq:euler}, we now advance the solution in time within the TN framework using the Euler method. The state $\mathbf{u}_t$, initially mapped into MPS form as described in the previous section, is updated iteratively by applying the operator $\mathcal{L}$ through successive MPO--MPS contractions.

Each update involves both operator application and summation of MPSs. These operations generally increase the bond dimensions of the resulting MPS, which in turn raises memory usage and computational cost. To keep the simulation tractable, we perform an SVD-based truncation after every time step, reducing all internal bond dimensions to a prescribed maximum $\chi$. This truncation retains the dominant singular values and discards subdominant contributions, thereby preserving the essential structure of the evolving state while controlling complexity.

By repeating this update scheme for the desired number of time steps, we simulate viscous transport governed by the advective-diffusion equation efficiently in the compressed TN representation, without ever leaving the reduced space.

\paragraph{Step 4: Decoding}

\hspace{5mm}To extract physical information from the compressed representation, one may employ localized reconstruction techniques such as those proposed by Peddinti et al.~\cite{Peddinti2024}, which allow partial recovery of the state and thereby reduce unnecessary computational costs. Here, however, for simplicity and because of its prevalence in the literature, we focus on the full recovery of the spatial vector $\mathbf{u}_t \in \mathbb{R}^{2^L}$ under periodic boundary conditions. This procedure involves contracting all MPS tensors to reconstruct the full tensor $\mathcal{U}_{i_1 i_2 \ldots i_L}$ and then flattening it back into a vector,
\begin{equation}
    \mathbf{u}_j = \mathcal{U}_{i_1 i_2 \ldots i_L}\,\,.
\end{equation}

\begin{figure}
    \centering
    \includegraphics[width=1.0\linewidth]{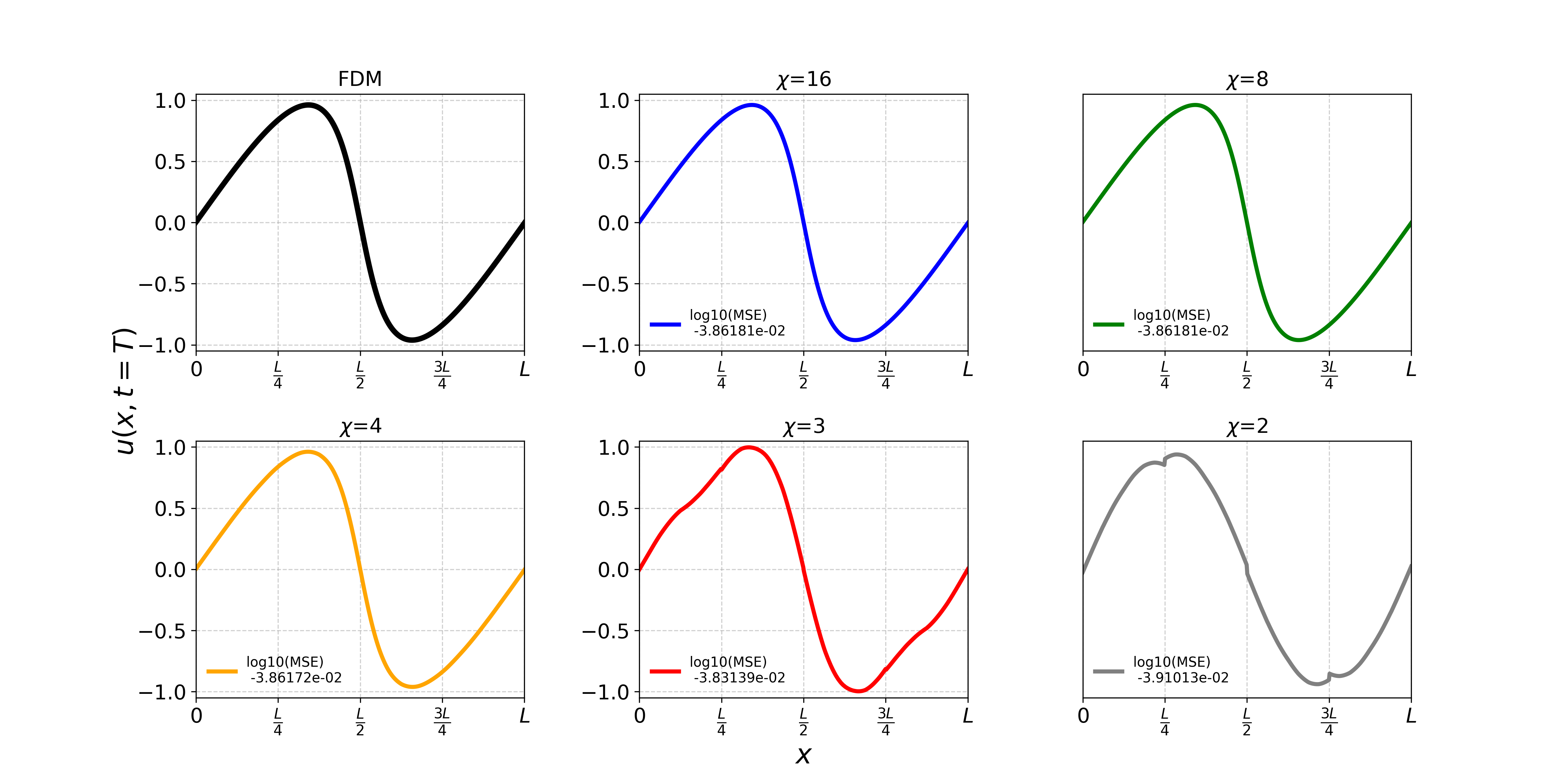}
    \caption{Final velocity field $u(x, T)$ for the 1D viscous Burgers equation, comparing the FDM reference solution with MPS-based approximations using varying bond dimensions $\chi$, where all approximations are using the same number of iterations (time discretization). The $\log_{10}$ of the Mean Squared Error (MSE) relative to the FDM solution is shown in each subplot.
}
    \label{fig:mps_example}
\end{figure}

Because the bond dimension $\chi$ is controlled during the time evolution, the reconstructed vector is an approximation of the exact solution. It preserves the dominant features of the dynamics while discarding less relevant correlations. This final step enables direct comparison between the TN-based solution and analytical or high-fidelity numerical references, as illustrated in Figure~\ref{fig:mps_example}.

This example highlights how variations in $\chi$ strongly affect solution quality. Even at low $\chi$, where compression is more severe, the qualitative shape of the solution still aligns with expectations. The MSE between the reference FDM result and the FDM+TN approximation is typically on the order of $10^{-2}$. Interestingly, the case $\chi=2$ exhibits a lower MSE than some higher-$\chi$ runs, yet its qualitative accuracy is worse, suggesting that this reduced error is coincidental rather than representative of a better approximation. Such observations emphasize the trade-off between computational efficiency (small $\chi$) and accuracy (large $\chi$) in TN-based simulations of nonlinear PDEs.

\begin{figure}[!h]
    \centering
    \includegraphics[width=\linewidth]{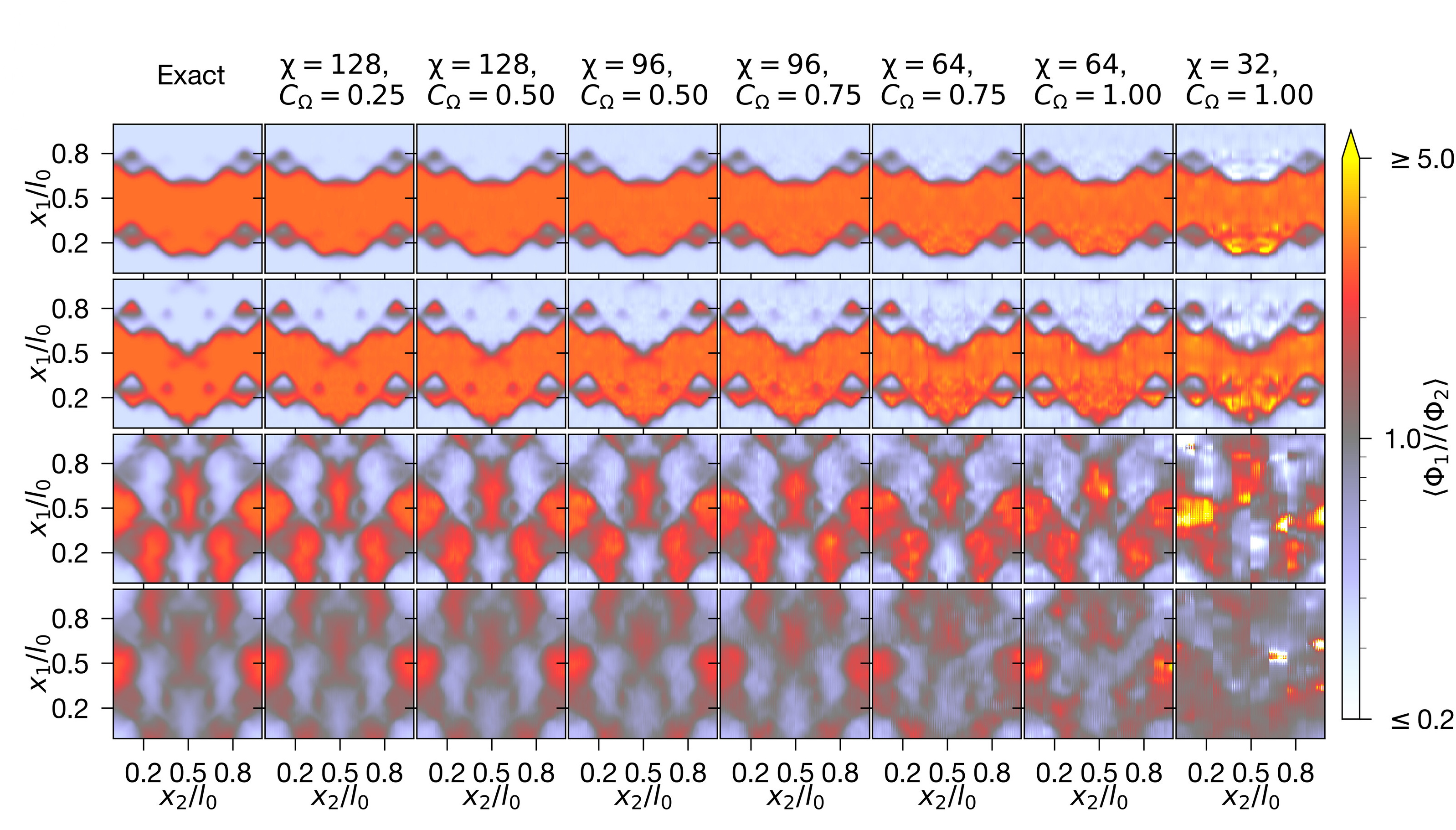}
    \caption{
    Comparison between exact solutions and tensor network (MPS) simulations for different values of the bond dimension $\chi$ and the mixing parameter $C_\Omega$. The figure demonstrates how simulation accuracy increases with higher $\chi$, particularly in regimes with strong subgrid-scale mixing. Even for moderate values of $\chi$, the MPS representation yields sufficiently accurate approximations of the turbulent flow PDF, capturing the key statistical features. 
    \textit{Source: Adapted from Gourianov et al. (2025), licensed under CC BY 4.0.}
    }
    \label{fig:gourianov}
\end{figure}

Although the Burgers equation serves as a relatively simple test case, more complex applications have been demonstrated. Figure~\ref{fig:gourianov} shows results from Gourianov et al., who applied TN methods to the challenging task of modeling the probability density function (PDF) of turbulent reactive flows. In their approach, the high-dimensional PDF is encoded as an MPS, enabling efficient time evolution of the Fokker–Planck equation. The figure illustrates how the bond dimension $\chi$ controls the expressive power of the MPS representation: smaller $\chi$ values reduce cost but degrade accuracy, while larger values better reproduce the exact solution. Remarkably, even moderate bond dimensions yield accurate approximations that capture key statistical features of the turbulent PDF, closely matching finite-difference benchmarks.

In summary, the decoding step illustrates both the strengths and limitations of TN methods. Their main advantage lies in scalability: by working entirely in reduced-rank form, simulations of otherwise intractable high-dimensional problems become feasible. When the underlying dynamics admit efficient low-rank representations, TN-based solvers can achieve substantial reductions in memory and runtime compared to conventional CFD approaches, as also reflected in the benchmarks summarized in Table~\ref{tab:tn_cfd}.

\section{Perspectives}
\label{sec:perspectives}

As discussed throughout this work, quantum computing represents a promising avenue for addressing the dimensionality challenges inherent in CFD, particularly those arising from the exponential growth of the Hilbert space. Among the most prominent proposals are VQAs, which exploit hybrid quantum-classical optimization schemes to approximate solutions with reduced resource requirements. Beyond VQAs, quantum architectures also enable richer exploration of parameter spaces in machine learning, motivating approaches that combine classical and quantum components. Within this context, QPINNs and HPINNs emerge as natural candidates, as they embed physical constraints directly into the learning process and may leverage quantum resources to improve parameter efficiency and accuracy. Such architectures have the potential to discover solutions beyond the reach of purely classical models. Nevertheless, a major challenge persists: in the current NISQ era, we lack both sufficiently robust qubits and a large enough number of them to tackle problems at the enormous spatial and temporal scales characteristic of real-world CFD, such as turbulent flows governed by the Navier-Stokes equations.

At present, studying turbulence in a fully quantum regime remains unfeasible and is best regarded as a long-term objective. On the other hand, significant progress has been made with quantum-inspired techniques, which exploit the representational efficiency of quantum computing concepts as compressed state-space formalisms. Recent studies have shown that such methods can mitigate the curse of dimensionality not only in quantum mechanics but also in CFD, achieving substantial gains in memory efficiency and computational speed relative to traditional methods such as FEM, FVM, and FDM. Importantly, TNs  do not replace these numerical methods; rather, they complement them, and have already shown promise even in turbulence-related problems.

In the near term, the most promising direction lies in the development of hybrid algorithms that combine TN-based compression with established numerical schemes, while also exploiting modern hardware accelerators (GPUs, TPUs) for efficient tensor operations. Even when implemented entirely on classical hardware, such approaches represent a transition toward quantum-inspired computing paradigms, offering a fertile ground for research until fault-tolerant quantum devices become available. Indeed, quantum-inspired methods are already applied in practice, including commercially for fluid dynamics problems~\cite{Terra1, Terra2, QIEO, Azure}.

Finally, it is worth emphasizing that quantum computing and TNs can be employed synergistically in two complementary ways. First, TNs can be used on classical hardware to efficiently simulate quantum computers with many qubits and deep circuits~\cite{Berezutskii2025}, enabling studies in quantum machine learning (QML) and variational quantum algorithms (VQAs) that would otherwise be infeasible. Second, TNs can serve as a preprocessing step for quantum computers themselves: rather than inserting the full CFD problem directly into a quantum circuit, the information can first be compressed via TNs and then mapped onto quantum hardware. This strategy reduces the required resources and allows larger-scale problems to be tackled. In this sense, TNs not only provide powerful tools on classical computers but also offer a natural structure that can be reused and adapted to future quantum architectures~\cite{Kornev2023}.

\section*{Statements and Declarations}

\bmhead{Funding}
This work was supported by the Coordenação de Aperfeiçoamento de Pessoal de Nível Superior (CAPES), Brazil - Finance Code 001, and by the Conselho Nacional de Desenvolvimento Científico e Tecnológico (CNPq), Brazil (Grant No.  409673/2022-6). EID also acknowledges the support of the Instituto Nacional de Ciência e Tecnologia de Informação Quântica (INCT-IQ). 

\bmhead{Author Contribution}
C.A.A. and V.L.O. prepared the results, generated all figures, and wrote the main manuscript text. J.P.L.C.S. and E.I.D. contributed with corrections and suggestions. All authors reviewed the manuscript.

\begin{appendices}

\end{appendices}

\bibliography{bibliography}

@article{Schollwock2011,
title = {The density-matrix renormalization group in the age of matrix product states},
journal = {Annals of Physics},
volume = {326},
number = {1},
pages = {96-192},
year = {2011},
note = {January 2011 Special Issue},
issn = {0003-4916},
doi = {https://doi.org/10.1016/j.aop.2010.09.012},
url = {https://www.sciencedirect.com/science/article/pii/S0003491610001752},
author = {Ulrich Schollwöck}
}

@phdthesis{gourianov2022a,
  edition = {},
  number = {},
  journal = {},
  booktitle = {},
  pages = {},
  publisher = {University of Oxford},
  school = {University of Oxford},
  title = {Exploiting the structure of turbulence with tensor networks},
  volume = {},
  author = {Gourianov, N},
  editor = {},
  year = {2022},
  series = {}
}

@article{clader2013preconditioned,
  title={Preconditioned quantum linear system algorithm},
  author={Clader, B David and Jacobs, Bryan C and Sprouse, Chad R},
  journal={Physical review letters},
  volume={110},
  number={25},
  pages={250504},
  year={2013},
  publisher={APS}
}

@article{cao2013quantum,
  title={Quantum algorithm and circuit design solving the Poisson equation},
  author={Cao, Yudong and Papageorgiou, Anargyros and Petras, Iasonas and Traub, Joseph and Kais, Sabre},
  journal={New Journal of Physics},
  volume={15},
  number={1},
  pages={013021},
  year={2013},
  publisher={IOP Publishing}
}

@article{montanaro2016quantum,
  title={Quantum algorithms and the finite element method},
  author={Montanaro, Ashley and Pallister, Sam},
  journal={Physical Review A},
  volume={93},
  number={3},
  pages={032324},
  year={2016},
  publisher={APS}
}

@article{costa2019quantum,
  title={Quantum algorithm for simulating the wave equation},
  author={Costa, Pedro CS and Jordan, Stephen and Ostrander, Aaron},
  journal={Physical Review A},
  volume={99},
  number={1},
  pages={012323},
  year={2019},
  publisher={APS}
}

@article{harrow2009quantum,
  title={Quantum algorithm for linear systems of equations},
  author={Harrow, Aram W and Hassidim, Avinatan and Lloyd, Seth},
  journal={Physical review letters},
  volume={103},
  number={15},
  pages={150502},
  year={2009},
  publisher={APS}
}

@article{pollachini2021hybrid,
  title={Hybrid classical-quantum approach to solve the heat equation using quantum annealers},
  author={Pollachini, Giovani G and Salazar, Juan PLC and G{\'o}es, Caio BD and Maciel, Thiago O and Duzzioni, Eduardo I},
  journal={Physical Review A},
  volume={104},
  number={3},
  pages={032426},
  year={2021},
  publisher={APS}
}

@misc{Kornev2023,
      title={Numerical solution of the incompressible Navier-Stokes equations for chemical mixers via quantum-inspired Tensor Train Finite Element Method}, 
      author={Egor Kornev and Sergey Dolgov and Karan Pinto and Markus Pflitsch and Michael Perelshtein and Artem Melnikov},
      year={2023},
      eprint={2305.10784},
      archivePrefix={arXiv},
      primaryClass={physics.flu-dyn},
      url={https://arxiv.org/abs/2305.10784}, 
}

@article{
Nikita2025,
author = {Nikita Gourianov  and Peyman Givi  and Dieter Jaksch  and Stephen B. Pope },
title = {Tensor networks enable the calculation of turbulence probability
                    distributions},
journal = {Science Advances},
volume = {11},
number = {5},
pages = {eads5990},
year = {2025},
doi = {10.1126/sciadv.ads5990},
URL = {https://www.science.org/doi/abs/10.1126/sciadv.ads5990},
eprint = {https://www.science.org/doi/pdf/10.1126/sciadv.ads5990}}

@misc{Pisoni2025,
      title={Compression, simulation, and synthesis of turbulent flows with tensor trains}, 
      author={Stefano Pisoni and Raghavendra Dheeraj Peddinti and Egor Tiunov and Siddhartha E. Guzman and Leandro Aolita},
      year={2025},
      eprint={2506.05477},
      archivePrefix={arXiv},
      primaryClass={physics.flu-dyn},
      url={https://arxiv.org/abs/2506.05477}, 
}

@article{Orus2014,
title = {A practical introduction to tensor networks: Matrix product states and projected entangled pair states},
journal = {Annals of Physics},
volume = {349},
pages = {117-158},
year = {2014},
issn = {0003-4916},
doi = {https://doi.org/10.1016/j.aop.2014.06.013},
url = {https://www.sciencedirect.com/science/article/pii/S0003491614001596},
author = {Román Orús}
}

@article{Kiffer2023,
  title = {Tensor network reduced order models for wall-bounded flows},
  author = {Kiffner, Martin and Jaksch, Dieter},
  journal = {Phys. Rev. Fluids},
  volume = {8},
  issue = {12},
  pages = {124101},
  numpages = {20},
  year = {2023},
  month = {Dec},
  publisher = {American Physical Society},
  doi = {10.1103/PhysRevFluids.8.124101},
  url = {https://link.aps.org/doi/10.1103/PhysRevFluids.8.124101}
}

@Article{Peddinti2024,
author={Peddinti, Raghavendra Dheeraj
and Pisoni, Stefano
and Marini, Alessandro
and Lott, Philippe
and Argentieri, Henrique
and Tiunov, Egor
and Aolita, Leandro},
title={Quantum-inspired framework for computational fluid dynamics},
journal={Communications Physics},
year={2024},
month={Apr},
day={27},
volume={7},
number={1},
pages={135},
issn={2399-3650},
doi={10.1038/s42005-024-01623-8},
url={https://doi.org/10.1038/s42005-024-01623-8}
}

@article{Holscher2025,
  title = {Quantum-inspired fluid simulation of two-dimensional turbulence with GPU acceleration},
  author = {H\"olscher, Leonhard and Rao, Pooja and M\"uller, Lukas and Klepsch, Johannes and Luckow, Andre and Stollenwerk, Tobias and Wilhelm, Frank K.},
  journal = {Phys. Rev. Res.},
  volume = {7},
  issue = {1},
  pages = {013112},
  numpages = {17},
  year = {2025},
  month = {Jan},
  publisher = {American Physical Society},
  doi = {10.1103/PhysRevResearch.7.013112},
  url = {https://link.aps.org/doi/10.1103/PhysRevResearch.7.013112}
}

@article{Danis2025,
title = {Tensor-train WENO scheme for compressible flows},
journal = {Journal of Computational Physics},
volume = {529},
pages = {113891},
year = {2025},
issn = {0021-9991},
doi = {https://doi.org/10.1016/j.jcp.2025.113891},
url = {https://www.sciencedirect.com/science/article/pii/S0021999125001743},
author = {M. Engin Danis and Duc Truong and Ismael Boureima and Oleg Korobkin and Kim {\O}. Rasmussen and Boian S. Alexandrov}
}

@article{Walton2025,
title = {The tensor-train stochastic finite volume method for uncertainty quantification},
journal = {Journal of Computational Physics},
volume = {538},
pages = {114192},
year = {2025},
issn = {0021-9991},
doi = {https://doi.org/10.1016/j.jcp.2025.114192},
url = {https://www.sciencedirect.com/science/article/pii/S0021999125004759},
author = {Steven Walton and Svetlana Tokareva and Gianmarco Manzini}
}

@misc{tensornetwork,
  author       = {{TensorNetwork Project}},
  title        = {TensorNetwork},
  year         = {n.d.},
  howpublished = {\url{https://tensornetwork.org}},
  note         = {Acesso em 23 de junho de 2025}
}

@misc{Vanhulst2025,
      title={Quantum-Inspired Tensor-Network Fractional-Step Method for Incompressible Flow in Curvilinear Coordinates}, 
      author={Nis-Luca van Hülst and Pia Siegl and Paul Over and Sergio Bengoechea and Tomohiro Hashizume and Mario Guillaume Cecile and Thomas Rung and Dieter Jaksch},
      year={2025},
      eprint={2507.05222},
      archivePrefix={arXiv},
      primaryClass={physics.flu-dyn},
      url={https://arxiv.org/abs/2507.05222}, 
}

@inbook{Sinan2024,
author = {Sinan Demir and Ryan Johnson and Brian T. Bojko and Andrew T. Corrigan and David A. Kessler and Pierson Guthrey and Jason Burmark and Samuel Irving},
title = {Three-Dimensional Compressible Chemically Reacting Computational Fluid Dynamics with Tensor Trains},
booktitle = {AIAA SCITECH 2024 Forum},
chapter = {},
pages = {},
doi = {10.2514/6.2024-1337},
URL = {https://arc.aiaa.org/doi/abs/10.2514/6.2024-1337},
eprint = {https://arc.aiaa.org/doi/pdf/10.2514/6.2024-1337}
}

@article{Ostlund1995,
  title = {Thermodynamic Limit of Density Matrix Renormalization},
  author = {\"Ostlund, Stellan and Rommer, Stefan},
  journal = {Phys. Rev. Lett.},
  volume = {75},
  issue = {19},
  pages = {3537--3540},
  numpages = {0},
  year = {1995},
  month = {Nov},
  publisher = {American Physical Society},
  doi = {10.1103/PhysRevLett.75.3537},
  url = {https://link.aps.org/doi/10.1103/PhysRevLett.75.3537}
}

@misc{Verstraete2004,
      title={Renormalization algorithms for Quantum-Many Body Systems in two and higher dimensions}, 
      author={F. Verstraete and J. I. Cirac},
      year={2004},
      eprint={cond-mat/0407066},
      archivePrefix={arXiv},
      primaryClass={cond-mat.str-el},
      url={https://arxiv.org/abs/cond-mat/0407066}, 
}

@article{Vidal2006,
  title = {Classical simulation of quantum many-body systems with a tree tensor network},
  author = {Shi, Y.-Y. and Duan, L.-M. and Vidal, G.},
  journal = {Phys. Rev. A},
  volume = {74},
  issue = {2},
  pages = {022320},
  numpages = {4},
  year = {2006},
  month = {Aug},
  publisher = {American Physical Society},
  doi = {10.1103/PhysRevA.74.022320},
  url = {https://link.aps.org/doi/10.1103/PhysRevA.74.022320}
}

@article{Vidal2008,
  title = {Class of Quantum Many-Body States That Can Be Efficiently Simulated},
  author = {Vidal, G.},
  journal = {Phys. Rev. Lett.},
  volume = {101},
  issue = {11},
  pages = {110501},
  numpages = {4},
  year = {2008},
  month = {Sep},
  publisher = {American Physical Society},
  doi = {10.1103/PhysRevLett.101.110501},
  url = {https://link.aps.org/doi/10.1103/PhysRevLett.101.110501}
}

@Article{Khoromskij2011,
author={Khoromskij, Boris N.},
title={O(dlog{\thinspace}N)-Quantics Approximation of N-d Tensors in High-Dimensional Numerical Modeling},
journal={Constructive Approximation},
year={2011},
month={Oct},
day={01},
volume={34},
number={2},
pages={257-280},
issn={1432-0940},
doi={10.1007/s00365-011-9131-1},
url={https://doi.org/10.1007/s00365-011-9131-1}
}

@Article{Dolgov2012,
author={Dolgov, Sergey
and Khoromskij, Boris
and Savostyanov, Dmitry},
title={Superfast Fourier Transform Using QTT Approximation},
journal={Journal of Fourier Analysis and Applications},
year={2012},
month={Oct},
day={01},
volume={18},
number={5},
pages={915-953},
issn={1531-5851},
doi={10.1007/s00041-012-9227-4},
url={https://doi.org/10.1007/s00041-012-9227-4}
}

@Article{Oseledets2013,
author={Oseledets, I. V.},
title={Constructive Representation of Functions in Low-Rank Tensor Formats},
journal={Constructive Approximation},
year={2013},
month={Feb},
day={01},
volume={37},
number={1},
pages={1-18},
issn={1432-0940},
doi={10.1007/s00365-012-9175-x},
url={https://doi.org/10.1007/s00365-012-9175-x}
}

@article{Osedelets2010,
title = {TT-cross approximation for multidimensional arrays},
journal = {Linear Algebra and its Applications},
volume = {432},
number = {1},
pages = {70-88},
year = {2010},
issn = {0024-3795},
doi = {https://doi.org/10.1016/j.laa.2009.07.024},
url = {https://www.sciencedirect.com/science/article/pii/S0024379509003747},
author = {Ivan Oseledets and Eugene Tyrtyshnikov},
}

@article{Oseledets2011,
author = {Oseledets, I. V.},
title = {Tensor-Train Decomposition},
journal = {SIAM Journal on Scientific Computing},
volume = {33},
number = {5},
pages = {2295-2317},
year = {2011},
doi = {10.1137/090752286},
URL = { 
    https://doi.org/10.1137/090752286
},
eprint = { 
    https://doi.org/10.1137/090752286
}
}

@misc{Terra1,
  author       = {Terra Quantum and Evonik},
  title        = {Terra Quantum and Evonik pave the way for advanced hybrid quantum solutions in CFD},
  howpublished = {\url{https://terraquantum.swiss/news/terra-quantum-and-evonik-pave-the-way-for-advanced-hybrid-quantum-solutions-in-cfd}},
  note         = {Accessed: 2025-09-06},
  year         = {2023}
}

@misc{Terra2,
  author       = {Terra Quantum and Evonik},
  title        = {Terra Quantum and Evonik successfully complete geometry optimization project},
  howpublished = {\url{https://terraquantum.swiss/news/terra-quantum-evonik-successfully-complete-geometry-optimization-project}},
  note         = {Accessed: 2025-09-06},
  year         = {2023}
}

@Article{Berezutskii2025,
author={Berezutskii, Aleksandr
and Liu, Minzhao
and Acharya, Atithi
and Ellerbrock, Roman
and Gray, Johnnie
and Haghshenas, Reza
and He, Zichang
and Khan, Abid
and Kuzmin, Viacheslav
and Lyakh, Dmitry
and Lykov, Danylo
and Mandr{\`a}, Salvatore
and Mansell, Christopher
and Melnikov, Alexey
and Melnikov, Artem
and Mironov, Vladimir
and Morozov, Dmitry
and Neukart, Florian
and Nocera, Alberto
and Perlin, Michael A.
and Perelshtein, Michael
and Steinberg, Matthew
and Shaydulin, Ruslan
and Villalonga, Benjamin
and Pflitsch, Markus
and Pistoia, Marco
and Vinokur, Valerii
and Alexeev, Yuri},
title={Tensor networks for quantum computing},
journal={Nature Reviews Physics},
year={2025},
month={Jul},
day={30},
issn={2522-5820},
doi={10.1038/s42254-025-00853-1},
url={https://doi.org/10.1038/s42254-025-00853-1}
}

@misc{QIEO,
  author       = {BQP Simulations},
  title        = {The Future of Aerospace with Quantum-Inspired Simulation},
  howpublished = {\url{https://www.bqpsim.com/blogs/future-of-aerospace-with-quantum-inspired-simulation}},
  note         = {Accessed: 2025-09-06},
  year         = {2024}
}

@book{Pope2000,
  author    = {Pope, Stephen B.},
  title     = {Turbulent Flows},
  publisher = {Cambridge University Press},
  year      = {2000},
  address   = {Cambridge; New York},
  isbn      = {978-0521598866},
  pages     = {xxxiv + 771},
}

@book{Ferziger2012,
  title     = {Computational Methods for Fluid Dynamics},
  author    = {Ferziger, J. H. and Peri{\'c}, M.},
  year      = {2012},
  publisher = {Springer},
  address   = {Berlin, Heidelberg},
  isbn      = {9783642560262}
}

@article{Larocca2025,
  author = {Larocca, Martín and Thanasilp, Supanut and Wang, Samson and
    Sharma, Kunal and Biamonte, Jacob and Coles, Patrick J. and Cincio,
    Lukasz and McClean, Jarrod R. and Holmes, Zoë and Cerezo, M.},
  title = {Barren Plateaus in Variational Quantum Computing},
  journal = {Nature Reviews Physics},
  volume = {7},
  number = {4},
  pages = {174-189},
  year = {2025},
  url = {https://doi.org/10.1038/s42254-025-00813-9}
}

@misc{Azure,
	author = {azure-quantum-content},
	title = {{W}hat is {A}zure {Q}uantum? - {A}zure {Q}uantum  --- learn.microsoft.com},
	howpublished = {\url{https://learn.microsoft.com/en-us/azure/quantum/overview-azure-quantum}},
	year = {},
	note = {[Accessed 06-09-2025]},
}

@Article{Qi2023,
author={Qi, Han
and Wang, Lei
and Zhu, Hongsheng
and Gani, Abdullah
and Gong, Changqing},
title={The barren plateaus of quantum neural networks: review, taxonomy and trends},
journal={Quantum Information Processing},
year={2023},
month={Dec},
day={11},
volume={22},
number={12},
pages={435},
issn={1573-1332},
doi={10.1007/s11128-023-04188-7},
url={https://doi.org/10.1007/s11128-023-04188-7}
}

@article{Verstraete2008,
author = {F. Verstraete and V. Murg and J.I. Cirac and},
title = {Matrix product states, projected entangled pair states, and variational renormalization group methods for quantum spin systems},
journal = {Advances in Physics},
volume = {57},
number = {2},
pages = {143--224},
year = {2008},
publisher = {Taylor \& Francis},
doi = {10.1080/14789940801912366},
URL = {     https://doi.org/10.1080/14789940801912366
},
eprint = { https://doi.org/10.1080/14789940801912366
}
}

@misc{Dubois2025,
      title={High-dimensional stochastic finite volumes using the tensor train format}, 
      author={Juliette Dubois and Michael Herty and Siegfried Müller},
      year={2025},
      eprint={2502.04868},
      archivePrefix={arXiv},
      primaryClass={math.NA},
      url={https://arxiv.org/abs/2502.04868}, 
}

@misc{Danis2024,
      title={High-order Tensor-Train Finite Volume Method for Shallow Water Equations}, 
      author={Mustafa Engin Danis and Duc P. Truong and Derek DeSantis and Mark Petersen and Kim O. Rasmussen and Boian S. Alexandrov},
      year={2024},
      eprint={2408.03483},
      archivePrefix={arXiv},
      primaryClass={math.NA},
      url={https://arxiv.org/abs/2408.03483}, 
}

@article{Kazeev2012,
author = {Kazeev, Vladimir A. and Khoromskij, Boris N.},
title = {Low-Rank Explicit QTT Representation of the Laplace Operator and Its Inverse},
journal = {SIAM Journal on Matrix Analysis and Applications},
volume = {33},
number = {3},
pages = {742-758},
year = {2012},
doi = {10.1137/100820479},
URL = { https://doi.org/10.1137/100820479
},
eprint = { 
    
        https://doi.org/10.1137/100820479
}
}

@article{Garcia2021,
  doi = {10.22331/q-2021-04-15-431},
  url = {https://doi.org/10.22331/q-2021-04-15-431},
  title = {Quantum-inspired algorithms for multivariate analysis: from interpolation to partial differential equations},
  author = {Garc{\'{i}}a-Ripoll, Juan Jos{\'{e}}},
  journal = {{Quantum}},
  issn = {2521-327X},
  publisher = {{Verein zur F{\"{o}}rderung des Open Access Publizierens in den Quantenwissenschaften}},
  volume = {5},
  pages = {431},
  month = apr,
  year = {2021}
}

@ARTICLE{Newman2004,
AUTHOR={Newman, Elizabeth  and Horesh, Lior  and Avron, Haim  and Kilmer, Misha E. },
TITLE={Stable tensor neural networks for efficient deep learning},
JOURNAL={Frontiers in Big Data},
VOLUME={Volume 7 - 2024},
YEAR={2024},
URL={https://www.frontiersin.org/journals/big-data/articles/10.3389/fdata.2024.1363978},
DOI={10.3389/fdata.2024.1363978},
ISSN={2624-909X}}

@misc{Wang2025,
      title={Tensor Networks Meet Neural Networks: A Survey and Future Perspectives}, 
      author={Maolin Wang and Yu Pan and Zenglin Xu and Guangxi Li and Xiangli Yang and Danilo Mandic and Andrzej Cichocki},
      year={2025},
      eprint={2302.09019},
      archivePrefix={arXiv},
      primaryClass={cs.LG},
      url={https://arxiv.org/abs/2302.09019}, 
}

@misc{Stoudenmire2017,
      title={Supervised Learning with Quantum-Inspired Tensor Networks}, 
      author={E. Miles Stoudenmire and David J. Schwab},
      year={2017},
      eprint={1605.05775},
      archivePrefix={arXiv},
      primaryClass={stat.ML},
      url={https://arxiv.org/abs/1605.05775}, 
}

@article{Liu2023,
  title = {Tensor networks for unsupervised machine learning},
  author = {Liu, Jing and Li, Sujie and Zhang, Jiang and Zhang, Pan},
  journal = {Phys. Rev. E},
  volume = {107},
  issue = {1},
  pages = {L012103},
  numpages = {6},
  year = {2023},
  month = {Jan},
  publisher = {American Physical Society},
  doi = {10.1103/PhysRevE.107.L012103},
  url = {https://link.aps.org/doi/10.1103/PhysRevE.107.L012103}
}

@article{scherer2017concrete,
  title={Concrete resource analysis of the quantum linear-system algorithm used to compute the electromagnetic scattering cross section of a 2D target},
  author={Scherer, Artur and Valiron, Beno{\^\i}t and Mau, Siun-Chuon and Alexander, Scott and Van den Berg, Eric and Chapuran, Thomas E},
  journal={Quantum Information Processing},
  volume={16},
  number={3},
  pages={1--65},
  year={2017},
  publisher={Springer}
}

@article{linden2022quantum,
  title={Quantum vs. classical algorithms for solving the heat equation},
  author={Linden, Noah and Montanaro, Ashley and Shao, Changpeng},
  journal={Communications in Mathematical Physics},
  volume={395},
  number={2},
  pages={601--641},
  year={2022},
  publisher={Springer}
}

@book{Burden,
author = {Burden, Annette and Burden, Richard and Faires, J.},
year = {2016},
month = {01},
pages = {},
title = {Numerical Analysis, 10th ed.},
isbn = {1305253663},
doi = {10.13140/2.1.4830.2406}
}

@article{Rieser2023,
   title={Tensor networks for quantum machine learning},
   volume={479},
   ISSN={1471-2946},
   url={http://dx.doi.org/10.1098/rspa.2023.0218},
   DOI={10.1098/rspa.2023.0218},
   number={2275},
   journal={Proceedings of the Royal Society A: Mathematical, Physical and Engineering Sciences},
   publisher={The Royal Society},
   author={Rieser, Hans-Martin and Köster, Frank and Raulf, Arne Peter},
   year={2023},
   month=jul }

@article{Jaksch2023,
author = {Jaksch, Dieter and Givi, Peyman and Daley, Andrew J. and Rung, Thomas},
title = {Variational Quantum Algorithms for Computational Fluid Dynamics},
journal = {AIAA Journal},
volume = {61},
number = {5},
pages = {1885-1894},
year = {2023},
doi = {10.2514/1.J062426},

URL = { 
    
        https://doi.org/10.2514/1.J062426
    
    

},
eprint = { 
    
        https://doi.org/10.2514/1.J062426
    
    

}
}

@book{BURGERS1948171,
title = {A Mathematical Model Illustrating the Theory of Turbulence},
editor = {Richard {Von Mises} and Theodore {Von Kármán}},
series = {Advances in Applied Mechanics},
publisher = {Elsevier},
volume = {1},
pages = {171-199},
year = {1948},
issn = {0065-2156},
doi = {https://doi.org/10.1016/S0065-2156(08)70100-5},
author = {J.M. Burgers}
}

@misc{Bosco2024,
      title={Demonstration of Scalability and Accuracy of Variational Quantum Linear Solver for Computational Fluid Dynamics}, 
      author={Ferdin Sagai Don Bosco and Dhamotharan S and Rut Lineswala and Abhishek Chopra},
      year={2024},
      eprint={2409.03241},
      archivePrefix={arXiv},
      primaryClass={physics.flu-dyn},
      url={https://arxiv.org/abs/2409.03241}, 
}

@article{Au2025,
   title={Quantum smoothed particle hydrodynamics algorithm inspired by quantum walks},
   volume={37},
   ISSN={1089-7666},
   url={http://dx.doi.org/10.1063/5.0268240},
   DOI={10.1063/5.0268240},
   number={5},
   journal={Physics of Fluids},
   publisher={AIP Publishing},
   author={Au-Yeung, R. and Kendon, V. M. and Lind, S. J.},
   year={2025},
   month=may }

@article{li_fourier_2021,
	title = {Fourier {Neural} {Operator} for {Parametric} {Partial} {Differential} {Equations}},
	url = {http://arxiv.org/abs/2010.08895},
	language = {en},
	urldate = {2021-09-27},
	journal = {arXiv:2010.08895 [cs, math]},
	author = {Li, Zongyi and Kovachki, Nikola and Azizzadenesheli, Kamyar and Liu, Burigede and Bhattacharya, Kaushik and Stuart, Andrew and Anandkumar, Anima},
	month = may,
	year = {2021},
	note = {arXiv: 2010.08895},
}

@book{roberts_principles_2022,
	title = {The {Principles} of {Deep} {Learning} {Theory}},
	url = {http://arxiv.org/abs/2106.10165},
	urldate = {2025-07-24},
	author = {Roberts, Daniel A. and Yaida, Sho and Hanin, Boris},
	month = may,
	year = {2022},
	doi = {10.1017/9781009023405},
	note = {arXiv:2106.10165 [cs]},
}

@misc{mangini_variational_2023,
	title = {Variational quantum algorithms for machine learning: theory and applications},
	shorttitle = {Variational quantum algorithms for machine learning},
	url = {http://arxiv.org/abs/2306.09984},
	doi = {10.48550/arXiv.2306.09984},
	urldate = {2025-07-24},
	publisher = {arXiv},
	author = {Mangini, Stefano},
	month = jun,
	year = {2023},
	note = {arXiv:2306.09984 [quant-ph]},
}

@article{schuld_effect_2021,
	title = {The effect of data encoding on the expressive power of variational quantum machine learning models},
	volume = {103},
	issn = {2469-9926, 2469-9934},
	url = {http://arxiv.org/abs/2008.08605},
	doi = {10.1103/PhysRevA.103.032430},
	number = {3},
	urldate = {2025-07-24},
	journal = {Physical Review A},
	author = {Schuld, Maria and Sweke, Ryan and Meyer, Johannes Jakob},
	month = mar,
	year = {2021},
	note = {arXiv:2008.08605 [quant-ph]},
}

@article{rath_quantum_2024,
	title = {Quantum data encoding: a comparative analysis of classical-to-quantum mapping techniques and their impact on machine learning accuracy},
	volume = {11},
	copyright = {2024 The Author(s)},
	issn = {2196-0763},
	shorttitle = {Quantum data encoding},
	url = {https://epjquantumtechnology.springeropen.com/articles/10.1140/epjqt/s40507-024-00285-3},
	doi = {10.1140/epjqt/s40507-024-00285-3},
	language = {en},
	number = {1},
	urldate = {2025-07-24},
	journal = {EPJ Quantum Technology},
	author = {Rath, Minati and Date, Hema},
	month = dec,
	year = {2024},
	note = {Number: 1
Publisher: SpringerOpen},
	pages = {1--22},
}

@misc{du_quantum_2025,
	title = {Quantum {Machine} {Learning}: {A} {Hands}-on {Tutorial} for {Machine} {Learning} {Practitioners} and {Researchers}},
	shorttitle = {Quantum {Machine} {Learning}},
	url = {http://arxiv.org/abs/2502.01146},
	doi = {10.48550/arXiv.2502.01146},
	urldate = {2025-07-25},
	publisher = {arXiv},
	author = {Du, Yuxuan and Wang, Xinbiao and Guo, Naixu and Yu, Zhan and Qian, Yang and Zhang, Kaining and Hsieh, Min-Hsiu and Rebentrost, Patrick and Tao, Dacheng},
	month = feb,
	year = {2025},
	note = {arXiv:2502.01146 [quant-ph]},
}

@article{schuld_evaluating_2019,
	title = {Evaluating analytic gradients on quantum hardware},
	volume = {99},
	issn = {2469-9926, 2469-9934},
	url = {http://arxiv.org/abs/1811.11184},
	doi = {10.1103/PhysRevA.99.032331},
	number = {3},
	urldate = {2025-07-25},
	journal = {Physical Review A},
	author = {Schuld, Maria and Bergholm, Ville and Gogolin, Christian and Izaac, Josh and Killoran, Nathan},
	month = mar,
	year = {2019},
	note = {arXiv:1811.11184 [quant-ph]},
}

@article{lubasch_variational_2020,
	title = {Variational quantum algorithms for nonlinear problems},
	volume = {101},
	url = {https://link.aps.org/doi/10.1103/PhysRevA.101.010301},
	doi = {10.1103/PhysRevA.101.010301},
	number = {1},
	urldate = {2025-07-25},
	journal = {Physical Review A},
	author = {Lubasch, Michael and Joo, Jaewoo and Moinier, Pierre and Kiffner, Martin and Jaksch, Dieter},
	month = jan,
	year = {2020},
	note = {Publisher: American Physical Society},
	pages = {010301},
}

@article{jaksch_variational_2023,
	title = {Variational {Quantum} {Algorithms} for {Computational} {Fluid} {Dynamics}},
	volume = {61},
	issn = {0001-1452},
	url = {https://arc.aiaa.org/doi/10.2514/1.J062426},
	doi = {10.2514/1.J062426},
	number = {5},
	urldate = {2025-07-25},
	journal = {AIAA Journal},
	author = {Jaksch, Dieter and Givi, Peyman and Daley, Andrew J. and Rung, Thomas},
	month = may,
	year = {2023},
	note = {Publisher: American Institute of Aeronautics and Astronautics},
	pages = {1885--1894},
}

@article{lubasch_multigrid_2018,
	title = {Multigrid {Renormalization}},
	volume = {372},
	issn = {0021-9991},
	url = {http://arxiv.org/abs/1802.07259},
	doi = {10.1016/j.jcp.2018.06.065},
	urldate = {2025-07-25},
	journal = {Journal of Computational Physics},
	author = {Lubasch, Michael and Moinier, Pierre and Jaksch, Dieter},
	month = nov,
	year = {2018},
	note = {arXiv:1802.07259 [physics]},
	pages = {587--602},
}

@article{raissi_physics-informed_2019,
	title = {Physics-informed neural networks: {A} deep learning framework for solving forward and inverse problems involving nonlinear partial differential equations},
	volume = {378},
	issn = {0021-9991},
	shorttitle = {Physics-informed neural networks},
	url = {https://www.sciencedirect.com/science/article/pii/S0021999118307125},
	doi = {10.1016/j.jcp.2018.10.045},
	urldate = {2025-07-30},
	journal = {Journal of Computational Physics},
	author = {Raissi, M. and Perdikaris, P. and Karniadakis, G. E.},
	month = feb,
	year = {2019},
	pages = {686--707},
}

@article{cerezo_variational_2021,
	title = {Variational quantum algorithms},
	volume = {3},
	copyright = {2021 Springer Nature Limited},
	issn = {2522-5820},
	url = {https://www.nature.com/articles/s42254-021-00348-9},
	doi = {10.1038/s42254-021-00348-9},
	language = {en},
	number = {9},
	urldate = {2025-07-30},
	journal = {Nature Reviews Physics},
	author = {Cerezo, M. and Arrasmith, Andrew and Babbush, Ryan and Benjamin, Simon C. and Endo, Suguru and Fujii, Keisuke and McClean, Jarrod R. and Mitarai, Kosuke and Yuan, Xiao and Cincio, Lukasz and Coles, Patrick J.},
	month = sep,
	year = {2021},
	note = {Publisher: Nature Publishing Group},
	pages = {625--644},
}

@article{sweke_stochastic_2020,
	title = {Stochastic gradient descent for hybrid quantum-classical optimization},
	volume = {4},
	issn = {2521-327X},
	url = {http://arxiv.org/abs/1910.01155},
	doi = {10.22331/q-2020-08-31-314},
	urldate = {2025-07-30},
	journal = {Quantum},
	author = {Sweke, Ryan and Wilde, Frederik and Meyer, Johannes and Schuld, Maria and Faehrmann, Paul K. and Meynard-Piganeau, Barthélémy and Eisert, Jens},
	month = aug,
	year = {2020},
	note = {arXiv:1910.01155 [quant-ph]},
	pages = {314},
}

@article{gourianov_quantum_2022,
	title = {A {Quantum} {Inspired} {Approach} to {Exploit} {Turbulence} {Structures}},
	volume = {2},
	issn = {2662-8457},
	url = {http://arxiv.org/abs/2106.05782},
	doi = {10.1038/s43588-021-00181-1},
	number = {1},
	urldate = {2025-07-30},
	journal = {Nature Computational Science},
	author = {Gourianov, Nikita and Lubasch, Michael and Dolgov, Sergey and Berg, Quincy Y. van den and Babaee, Hessam and Givi, Peyman and Kiffner, Martin and Jaksch, Dieter},
	month = jan,
	year = {2022},
	note = {arXiv:2106.05782 [physics]},
	pages = {30--37},
}

@misc{crooks_gradients_2019,
	title = {Gradients of parameterized quantum gates using the parameter-shift rule and gate decomposition},
	url = {http://arxiv.org/abs/1905.13311},
	doi = {10.48550/arXiv.1905.13311},
	urldate = {2025-07-30},
	publisher = {arXiv},
	author = {Crooks, Gavin E.},
	month = may,
	year = {2019},
	note = {arXiv:1905.13311 [quant-ph]},
}

@article{hornik_multilayer_1989,
	title = {Multilayer feedforward networks are universal approximators},
	volume = {2},
	issn = {0893-6080},
	url = {https://www.sciencedirect.com/science/article/pii/0893608089900208},
	doi = {10.1016/0893-6080(89)90020-8},
	number = {5},
	urldate = {2025-07-30},
	journal = {Neural Networks},
	author = {Hornik, Kurt and Stinchcombe, Maxwell and White, Halbert},
	month = jan,
	year = {1989},
	pages = {359--366},
}

@misc{guerreschi_practical_2017,
	title = {Practical optimization for hybrid quantum-classical algorithms},
	url = {http://arxiv.org/abs/1701.01450},
	doi = {10.48550/arXiv.1701.01450},
	urldate = {2025-08-04},
	publisher = {arXiv},
	author = {Guerreschi, Gian Giacomo and Smelyanskiy, Mikhail},
	month = jan,
	year = {2017},
	note = {arXiv:1701.01450 [quant-ph]},
}

@book{luongo_chapter_nodate,
	title = {Chapter 2 {Quantum} computing and quantum algorithms {\textbar} {Quantum} algorithms for data analysis},
	url = {https://quantumalgorithms.org/chapter-intro.html#hadamard-test},
	urldate = {2025-08-11},
	author = {Luongo, Alessandro},
}

@misc{raissi_hidden_2018,
	title = {Hidden {Fluid} {Mechanics}: {A} {Navier}-{Stokes} {Informed} {Deep} {Learning} {Framework} for {Assimilating} {Flow} {Visualization} {Data}},
	shorttitle = {Hidden {Fluid} {Mechanics}},
	url = {http://arxiv.org/abs/1808.04327},
	doi = {10.48550/arXiv.1808.04327},
	urldate = {2025-08-29},
	publisher = {arXiv},
	author = {Raissi, Maziar and Yazdani, Alireza and Karniadakis, George Em},
	month = aug,
	year = {2018},
	note = {arXiv:1808.04327 [cs]},
}

@article{eivazi_physics-informed_2022,
	title = {Physics-informed neural networks for solving {Reynolds}-averaged {Navier}\${\textbackslash}unicode\{x2013\}\${Stokes} equations},
	volume = {34},
	issn = {1070-6631, 1089-7666},
	url = {http://arxiv.org/abs/2107.10711},
	doi = {10.1063/5.0095270},
	number = {7},
	urldate = {2025-08-29},
	journal = {Physics of Fluids},
	author = {Eivazi, Hamidreza and Tahani, Mojtaba and Schlatter, Philipp and Vinuesa, Ricardo},
	month = jul,
	year = {2022},
	note = {arXiv:2107.10711 [physics]},
	pages = {075117},
}

@misc{gazoulis_stability_2025,
	title = {On the {Stability} and {Convergence} of {Physics} {Informed} {Neural} {Networks}},
	url = {http://arxiv.org/abs/2308.05423},
	doi = {10.48550/arXiv.2308.05423},
	urldate = {2025-08-29},
	publisher = {arXiv},
	author = {Gazoulis, Dimitrios and Gkanis, Ioannis and Makridakis, Charalambos G.},
	month = jul,
	year = {2025},
	note = {arXiv:2308.05423 [math]},
}

@article{mangini_quantum_2021,
	title = {Quantum computing models for artificial neural networks},
	volume = {134},
	issn = {0295-5075, 1286-4854},
	url = {http://arxiv.org/abs/2102.03879},
	doi = {10.1209/0295-5075/134/10002},
	number = {1},
	urldate = {2025-08-29},
	journal = {Europhysics Letters},
	author = {Mangini, Stefano and Tacchino, Francesco and Gerace, Dario and Bajoni, Daniele and Macchiavello, Chiara},
	month = apr,
	year = {2021},
	note = {arXiv:2102.03879 [quant-ph]},
	pages = {10002},
}

@article{ranga_quantum_2024,
	title = {Quantum {Machine} {Learning}: {Exploring} the {Role} of {Data} {Encoding} {Techniques}, {Challenges}, and {Future} {Directions}},
	volume = {12},
	copyright = {http://creativecommons.org/licenses/by/3.0/},
	issn = {2227-7390},
	shorttitle = {Quantum {Machine} {Learning}},
	url = {https://www.mdpi.com/2227-7390/12/21/3318},
	doi = {10.3390/math12213318},
	language = {en},
	number = {21},
	urldate = {2025-08-29},
	journal = {Mathematics},
	author = {Ranga, Deepak and Rana, Aryan and Prajapat, Sunil and Kumar, Pankaj and Kumar, Kranti and Vasilakos, Athanasios V.},
	month = jan,
	year = {2024},
	note = {Publisher: Multidisciplinary Digital Publishing Institute},
	pages = {3318},
}

@article{goto_universal_2021,
	title = {Universal {Approximation} {Property} of {Quantum} {Machine} {Learning} {Models} in {Quantum}-{Enhanced} {Feature} {Spaces}},
	volume = {127},
	issn = {0031-9007, 1079-7114},
	url = {http://arxiv.org/abs/2009.00298},
	doi = {10.1103/PhysRevLett.127.090506},
	number = {9},
	urldate = {2025-08-30},
	journal = {Physical Review Letters},
	author = {Goto, Takahiro and Tran, Quoc Hoan and Nakajima, Kohei},
	month = aug,
	year = {2021},
	note = {arXiv:2009.00298 [quant-ph]},
	pages = {090506},
}

@article{perez-salinas_data_2020,
	title = {Data re-uploading for a universal quantum classifier},
	volume = {4},
	issn = {2521-327X},
	url = {http://arxiv.org/abs/1907.02085},
	doi = {10.22331/q-2020-02-06-226},
	urldate = {2025-08-30},
	journal = {Quantum},
	author = {Pérez-Salinas, Adrián and Cervera-Lierta, Alba and Gil-Fuster, Elies and Latorre, José I.},
	month = feb,
	year = {2020},
	note = {arXiv:1907.02085 [quant-ph]},
	pages = {226},
}

@incollection{al-gwaiz_fourier_2008,
	address = {London},
	title = {Fourier {Series}},
	isbn = {978-1-84628-972-9},
	url = {https://doi.org/10.1007/978-1-84628-972-9_3},
	language = {en},
	urldate = {2025-08-30},
	booktitle = {Sturm-{Liouville} {Theory} and its {Applications}},
	publisher = {Springer},
	editor = {Al-Gwaiz, M. A.},
	year = {2008},
	doi = {10.1007/978-1-84628-972-9_3},
	pages = {93--128},
}

@article{kyriienko_solving_2021,
	title = {Solving nonlinear differential equations with differentiable quantum circuits},
	volume = {103},
	url = {https://link.aps.org/doi/10.1103/PhysRevA.103.052416},
	doi = {10.1103/PhysRevA.103.052416},
	number = {5},
	urldate = {2025-08-28},
	journal = {Physical Review A},
	author = {Kyriienko, Oleksandr and Paine, Annie E. and Elfving, Vincent E.},
	month = may,
	year = {2021},
	note = {Publisher: American Physical Society},
	pages = {052416},
}

@article{trahan_quantum_2024,
	title = {Quantum {Physics}-{Informed} {Neural} {Networks}},
	volume = {26},
	copyright = {http://creativecommons.org/licenses/by/3.0/},
	issn = {1099-4300},
	url = {https://www.mdpi.com/1099-4300/26/8/649},
	doi = {10.3390/e26080649},
	language = {en},
	number = {8},
	urldate = {2025-08-28},
	journal = {Entropy},
	author = {Trahan, Corey and Loveland, Mark and Dent, Samuel},
	month = aug,
	year = {2024},
	note = {Publisher: Multidisciplinary Digital Publishing Institute},
	pages = {649},
}

@article{schuld_circuit-centric_2020,
	title = {Circuit-centric quantum classifiers},
	volume = {101},
	issn = {2469-9926, 2469-9934},
	url = {http://arxiv.org/abs/1804.00633},
	doi = {10.1103/PhysRevA.101.032308},
	number = {3},
	urldate = {2025-08-31},
	journal = {Physical Review A},
	author = {Schuld, Maria and Bocharov, Alex and Svore, Krysta and Wiebe, Nathan},
	month = mar,
	year = {2020},
	note = {arXiv:1804.00633 [quant-ph]},
	pages = {032308},
}

@misc{leong_hybrid_2025,
	title = {Hybrid {Quantum} {Physics}-informed {Neural} {Network}: {Towards} {Efficient} {Learning} of {High}-speed {Flows}},
	shorttitle = {Hybrid {Quantum} {Physics}-informed {Neural} {Network}},
	url = {http://arxiv.org/abs/2503.02202},
	doi = {10.48550/arXiv.2503.02202},
	urldate = {2025-08-31},
	publisher = {arXiv},
	author = {Leong, Fong Yew and Ewe, Wei-Bin and Quang, Tran Si Bui and Zhang, Zhongyuan and Khoo, Jun Yong},
	month = aug,
	year = {2025},
	note = {arXiv:2503.02202 [physics]},
}

@article{sedykh_hybrid_2024,
	title = {Hybrid quantum physics-informed neural networks for simulating computational fluid dynamics in complex shapes},
	volume = {5},
	issn = {2632-2153},
	url = {https://dx.doi.org/10.1088/2632-2153/ad43b2},
	doi = {10.1088/2632-2153/ad43b2},
	language = {en},
	number = {2},
	urldate = {2025-08-28},
	journal = {Machine Learning: Science and Technology},
	author = {Sedykh, Alexandr and Podapaka, Maninadh and Sagingalieva, Asel and Pinto, Karan and Pflitsch, Markus and Melnikov, Alexey},
	month = may,
	year = {2024},
	note = {Publisher: IOP Publishing},
	pages = {025045},
}

@misc{farea_qcpinn_2025,
	title = {{QCPINN}: {Quantum} {Classical} {Physics}-{Informed} {Neural} {Networks} for {Solving} {PDEs}},
	shorttitle = {{QCPINN}},
	url = {http://arxiv.org/abs/2503.16678},
	doi = {10.48550/arXiv.2503.16678},
	urldate = {2025-08-28},
	publisher = {arXiv},
	author = {Farea, Afrah and Khan, Saiful and Celebi, Mustafa Serdar},
	month = mar,
	year = {2025},
	note = {arXiv:2503.16678 [quant-ph]
version: 1},
}

@article{gaitan_finding_2020,
	title = {Finding flows of a {Navier}–{Stokes} fluid through quantum computing},
	volume = {6},
	copyright = {2020 This is a U.S. government work and not under copyright protection in the U.S.; foreign copyright protection may apply},
	issn = {2056-6387},
	url = {https://www.nature.com/articles/s41534-020-00291-0},
	doi = {10.1038/s41534-020-00291-0},
	language = {en},
	number = {1},
	urldate = {2025-08-31},
	journal = {npj Quantum Information},
	author = {Gaitan, Frank},
	month = jul,
	year = {2020},
	note = {Publisher: Nature Publishing Group},
	pages = {61},
}

@article{mao_physics-informed_2020,
	title = {Physics-informed neural networks for high-speed flows},
	volume = {360},
	issn = {0045-7825},
	url = {https://www.sciencedirect.com/science/article/pii/S0045782519306814},
	doi = {10.1016/j.cma.2019.112789},
	urldate = {2025-09-01},
	journal = {Computer Methods in Applied Mechanics and Engineering},
	author = {Mao, Zhiping and Jagtap, Ameya D. and Karniadakis, George Em},
	month = mar,
	year = {2020},
	pages = {112789},
}

@Article{vonLarcher2019,
author={von Larcher, Thomas
and Klein, Rupert},
title={On identification of self-similar characteristics using the Tensor Train decomposition method with application to channel turbulence flow},
journal={Theoretical and Computational Fluid Dynamics},
year={2019},
month={Apr},
day={01},
volume={33},
number={2},
pages={141-159},
issn={1432-2250},
doi={10.1007/s00162-019-00485-z},
url={https://doi.org/10.1007/s00162-019-00485-z}
}

@article{mari_estimating_2021,
	title = {Estimating the gradient and higher-order derivatives on quantum hardware},
	volume = {103},
	url = {https://link.aps.org/doi/10.1103/PhysRevA.103.012405},
	doi = {10.1103/PhysRevA.103.012405},
	number = {1},
	urldate = {2025-09-03},
	journal = {Physical Review A},
	author = {Mari, Andrea and Bromley, Thomas R. and Killoran, Nathan},
	month = jan,
	year = {2021},
	note = {Publisher: American Physical Society},
	pages = {012405},
}

@article{biamonte_quantum_2017,
	title = {Quantum machine learning},
	volume = {549},
	copyright = {2017 Macmillan Publishers Limited, part of Springer Nature. All rights reserved.},
	issn = {1476-4687},
	url = {https://www.nature.com/articles/nature23474},
	doi = {10.1038/nature23474},
	language = {en},
	number = {7671},
	urldate = {2025-09-04},
	journal = {Nature},
	author = {Biamonte, Jacob and Wittek, Peter and Pancotti, Nicola and Rebentrost, Patrick and Wiebe, Nathan and Lloyd, Seth},
	month = sep,
	year = {2017},
	note = {Publisher: Nature Publishing Group},
	pages = {195--202},
}

@misc{panichi_quantum_2025,
	title = {Quantum physics informed neural networks for multi-variable partial differential equations},
	url = {http://arxiv.org/abs/2503.12244},
	doi = {10.48550/arXiv.2503.12244},
	urldate = {2025-09-05},
	publisher = {arXiv},
	author = {Panichi, Giorgio and Corli, Sebastiano and Prati, Enrico},
	month = mar,
	year = {2025},
	note = {arXiv:2503.12244 [quant-ph]},
}

@article{markidis_physics-informed_2022,
	title = {On {Physics}-{Informed} {Neural} {Networks} for {Quantum} {Computers}},
	volume = {8},
	issn = {2297-4687},
	url = {http://arxiv.org/abs/2209.14754},
	doi = {10.3389/fams.2022.1036711},
	urldate = {2025-09-05},
	journal = {Frontiers in Applied Mathematics and Statistics},
	author = {Markidis, Stefano},
	month = oct,
	year = {2022},
	note = {arXiv:2209.14754 [quant-ph]},
	pages = {1036711},
}

@article{bangar_experimentally_2023,
	title = {Experimentally realizable continuous-variable quantum neural networks},
	volume = {108},
	url = {https://link.aps.org/doi/10.1103/PhysRevA.108.042414},
	doi = {10.1103/PhysRevA.108.042414},
	number = {4},
	urldate = {2025-09-05},
	journal = {Physical Review A},
	author = {Bangar, Shikha and Sunny, Leanto and Yeter-Aydeniz, KÃ¼bra and Siopsis, George},
	month = oct,
	year = {2023},
	note = {Publisher: American Physical Society},
	pages = {042414},
}

@article{killoran_continuous-variable_2019,
	title = {Continuous-variable quantum neural networks},
	volume = {1},
	url = {https://link.aps.org/doi/10.1103/PhysRevResearch.1.033063},
	doi = {10.1103/PhysRevResearch.1.033063},
	number = {3},
	urldate = {2025-09-05},
	journal = {Physical Review Research},
	author = {Killoran, Nathan and Bromley, Thomas R. and Arrazola, Juan Miguel and Schuld, Maria and Quesada, NicolÃ¡s and Lloyd, Seth},
	month = oct,
	year = {2019},
	note = {Publisher: American Physical Society},
	pages = {033063},
}

@misc{wang_recent_2024,
	title = {Recent {Advances} on {Machine} {Learning} for {Computational} {Fluid} {Dynamics}: {A} {Survey}},
	shorttitle = {Recent {Advances} on {Machine} {Learning} for {Computational} {Fluid} {Dynamics}},
	url = {http://arxiv.org/abs/2408.12171},
	doi = {10.48550/arXiv.2408.12171},
	urldate = {2025-09-07},
	publisher = {arXiv},
	author = {Wang, Haixin and Cao, Yadi and Huang, Zijie and Liu, Yuxuan and Hu, Peiyan and Luo, Xiao and Song, Zezheng and Zhao, Wanjia and Liu, Jilin and Sun, Jinan and Zhang, Shikun and Wei, Long and Wang, Yue and Wu, Tailin and Ma, Zhi-Ming and Sun, Yizhou},
	month = aug,
	year = {2024},
	note = {arXiv:2408.12171 [cs]},
}

@article{kochkov_machine_2021,
	title = {Machine learning accelerated computational fluid dynamics},
	volume = {118},
	issn = {0027-8424, 1091-6490},
	url = {http://arxiv.org/abs/2102.01010},
	doi = {10.1073/pnas.2101784118},
	number = {21},
	urldate = {2025-09-07},
	journal = {Proceedings of the National Academy of Sciences},
	author = {Kochkov, Dmitrii and Smith, Jamie A. and Alieva, Ayya and Wang, Qing and Brenner, Michael P. and Hoyer, Stephan},
	month = may,
	year = {2021},
	note = {arXiv:2102.01010 [physics]},
	pages = {e2101784118},
}

@article{jin_nsfnets_2021,
	title = {{NSFnets} ({Navier}-{Stokes} {Flow} nets): {Physics}-informed neural networks for the incompressible {Navier}-{Stokes} equations},
	volume = {426},
	issn = {00219991},
	shorttitle = {{NSFnets} ({Navier}-{Stokes} {Flow} nets)},
	url = {http://arxiv.org/abs/2003.06496},
	doi = {10.1016/j.jcp.2020.109951},
	urldate = {2025-09-07},
	journal = {Journal of Computational Physics},
	author = {Jin, Xiaowei and Cai, Shengze and Li, Hui and Karniadakis, George Em},
	month = feb,
	year = {2021},
	note = {arXiv:2003.06496 [physics]},
	pages = {109951},
}

@article{vlatko_quantum_arithmetic_1996,
  title = {Quantum networks for elementary arithmetic operations},
  author = {Vedral, Vlatko and Barenco, Adriano and Ekert, Artur},
  journal = {Phys. Rev. A},
  volume = {54},
  issue = {1},
  pages = {147--153},
  numpages = {0},
  year = {1996},
  month = {Jul},
  publisher = {American Physical Society},
  doi = {10.1103/PhysRevA.54.147},
  url = {https://link.aps.org/doi/10.1103/PhysRevA.54.147}
}

@article{cerezo_fidelity_2020,
	title = {Variational {Quantum} {Fidelity} {Estimation}},
	volume = {4},
	issn = {2521-327X},
	url = {http://arxiv.org/abs/1906.09253},
	doi = {10.22331/q-2020-03-26-248},
	urldate = {2025-09-30},
	journal = {Quantum},
	author = {Cerezo, M. and Poremba, Alexander and Cincio, Lukasz and Coles, Patrick J.},
	month = mar,
	year = {2020},
	note = {arXiv:1906.09253 [quant-ph]},
	pages = {248},
}

@article{romero_quantum_2017,
	title = {Quantum autoencoders for efficient compression of quantum data},
	volume = {2},
	issn = {2058-9565},
	url = {http://arxiv.org/abs/1612.02806},
	doi = {10.1088/2058-9565/aa8072},
	number = {4},
	urldate = {2025-09-30},
	journal = {Quantum Science and Technology},
	author = {Romero, Jonathan and Olson, Jonathan P. and Aspuru-Guzik, Alan},
	month = dec,
	year = {2017},
	note = {arXiv:1612.02806 [quant-ph]},
	pages = {045001},
}

@article{dallaire-demers_quantum_2018,
	title = {Quantum generative adversarial networks},
	volume = {98},
	issn = {2469-9926, 2469-9934},
	url = {http://arxiv.org/abs/1804.08641},
	doi = {10.1103/PhysRevA.98.012324},
	number = {1},
	urldate = {2025-09-30},
	journal = {Physical Review A},
	author = {Dallaire-Demers, Pierre-Luc and Killoran, Nathan},
	month = jul,
	year = {2018},
	note = {arXiv:1804.08641 [quant-ph]},
	pages = {012324},
}

@article{tan_variational_2021,
	title = {Variational quantum algorithms to estimate rank, quantum entropies, fidelity and {Fisher} information via purity minimization},
	volume = {3},
	issn = {2643-1564},
	url = {http://arxiv.org/abs/2103.15956},
	doi = {10.1103/PhysRevResearch.3.033251},
	number = {3},
	urldate = {2025-09-30},
	journal = {Physical Review Research},
	author = {Tan, Kok Chuan and Volkoff, Tyler},
	month = sep,
	year = {2021},
	note = {arXiv:2103.15956 [quant-ph]},
	pages = {033251},
}

@article{sim_expressibility_2019,
	title = {Expressibility and entangling capability of parameterized quantum circuits for hybrid quantum-classical algorithms},
	volume = {2},
	issn = {2511-9044, 2511-9044},
	url = {http://arxiv.org/abs/1905.10876},
	doi = {10.1002/qute.201900070},
	number = {12},
	urldate = {2025-09-30},
	journal = {Advanced Quantum Technologies},
	author = {Sim, Sukin and Johnson, Peter D. and Aspuru-Guzik, Alan},
	month = dec,
	year = {2019},
	note = {arXiv:1905.10876 [quant-ph]},
	pages = {1900070},
}

@incollection{Bhattacharyya21,
author = {Suvanjan Bhattacharyya and John P. Abraham and Lijing Cheng and John Gorman},
title = {Introductory Chapter: A Brief History of and Introduction to Computational Fluid Dynamics},
booktitle = {Applications of Computational Fluid Dynamics Simulation and Modeling},
publisher = {IntechOpen},
address = {London},
year = {2021},
editor = {Suvanjan Bhattacharyya},
chapter = {1},
doi = {10.5772/intechopen.97235},
url = {https://doi.org/10.5772/intechopen.97235}
}

@book{anderson1995cfd,
  title     = {Computational Fluid Dynamics: The Basics with Applications},
  author    = {Anderson, John D.},
  year      = {1995},
  publisher = {McGraw-Hill},
  address   = {New York}
}

@book{batchelor1967fluid,
  title     = {An Introduction to Fluid Dynamics},
  author    = {Batchelor, George K.},
  year      = {1967},
  publisher = {Cambridge University Press},
  address   = {Cambridge, UK}
}

@article{charney1955primitive,
  title   = {The use of the primitive equations of motion in numerical prediction},
  author  = {Charney, Jule G.},
  journal = {Tellus},
  volume  = {7},
  number  = {1},
  pages   = {22--26},
  year    = {1955},
  publisher = {Wiley Online Library},
  doi     = {10.3402/tellusa.v7i1.8796}
}

@book{hirsch2007numerical,
  title     = {Numerical Computation of Internal and External Flows},
  author    = {Hirsch, Charles},
  edition   = {2},
  year      = {2007},
  publisher = {Elsevier},
  address   = {Oxford, UK}
}

@book{leveque2002finite,
  title     = {Finite Volume Methods for Hyperbolic Problems},
  author    = {LeVeque, Randall J.},
  year      = {2002},
  publisher = {Cambridge University Press},
  address   = {Cambridge, UK},
  doi       = {10.1017/CBO9780511791253}
}

@book{lomax1976fundamentals,
  title     = {Fundamentals of Computational Fluid Dynamics},
  author    = {Lomax, Harvey and Pulliam, Thomas H. and Zingg, David W.},
  year      = {1976},
  publisher = {NASA Reference Publication},
  address   = {Washington, DC}
}

@article{spalart2009des,
  title   = {Detached-eddy simulation},
  author  = {Spalart, Philippe R.},
  journal = {Annual Review of Fluid Mechanics},
  volume  = {41},
  pages   = {181--202},
  year    = {2009},
  publisher = {Annual Reviews},
  doi     = {10.1146/annurev.fluid.010908.165130}
}

@ARTICLE{stone_et_al_1992,
       author = {{Stone}, James M. and {Norman}, Michael L.},
        title = "{ZEUS-2D: A Radiation Magnetohydrodynamics Code for Astrophysical Flows in Two Space Dimensions. I. The Hydrodynamic Algorithms and Tests}",
      journal = {Astrophysical Journal Supplement},
         year = 1992,
        month = jun,
       volume = {80},
        pages = {753},
          doi = {10.1086/191680},
       adsurl = {https://ui.adsabs.harvard.edu/abs/1992ApJS...80..753S}
}

@article{taebi2024computational,
  title        = {Computational Fluid Dynamics in Medicine and Biology},
  author       = {Taebi, Amirtahà},
  journal      = {Bioengineering},
  year         = {2024},
  volume       = {11},
  number       = {11},
  pages        = {1168},
  doi          = {10.3390/bioengineering11111168},
  pmcid        = {PMC11591807},
  note         = {Editorial, Special Issue on CFD in medicine and biology}
}

@ARTICLE{Nishio_et_al_2025,
       author = {{Nishio}, Erika and {Tomida}, Kengo and {Kudoh}, Yuki and {Kimura}, Shigeo S.},
        title = "{Formation and Early Evolution of Protoplanetary Disks under Nonuniform Cosmic-Ray Ionization}",
      journal = {The Astrophysical Journal},
         year = 2025,
        month = jul,
       volume = {988},
       number = {1},
          eid = {56},
        pages = {56},
          doi = {10.3847/1538-4357/addbe2},
archivePrefix = {arXiv},
       eprint = {2505.09231},
 primaryClass = {astro-ph.SR},
       adsurl = {https://ui.adsabs.harvard.edu/abs/2025ApJ...988...56N}
}

@misc{streif_comparison_2019,
	title = {Comparison of {QAOA} with {Quantum} and {Simulated} {Annealing}},
	url = {http://arxiv.org/abs/1901.01903},
	doi = {10.48550/arXiv.1901.01903},
	urldate = {2025-09-30},
	publisher = {arXiv},
	author = {Streif, Michael and Leib, Martin},
	month = jan,
	year = {2019},
	note = {arXiv:1901.01903 [quant-ph]},
}

@article{peruzzo_variational_2014,
	title = {A variational eigenvalue solver on a quantum processor},
	volume = {5},
	issn = {2041-1723},
	url = {http://arxiv.org/abs/1304.3061},
	doi = {10.1038/ncomms5213},
	number = {1},
	urldate = {2025-09-30},
	journal = {Nature Communications},
	author = {Peruzzo, Alberto and McClean, Jarrod and Shadbolt, Peter and Yung, Man-Hong and Zhou, Xiao-Qi and Love, Peter J. and Aspuru-Guzik, Alán and O'Brien, Jeremy L.},
	month = jul,
	year = {2014},
	note = {arXiv:1304.3061 [quant-ph]},
	pages = {4213},
}

@misc{verdon_quantum_2019,
	title = {A quantum algorithm to train neural networks using low-depth circuits},
	url = {http://arxiv.org/abs/1712.05304},
	doi = {10.48550/arXiv.1712.05304},
	urldate = {2025-09-30},
	publisher = {arXiv},
	author = {Verdon, Guillaume and Broughton, Michael and Biamonte, Jacob},
	month = aug,
	year = {2019},
	note = {arXiv:1712.05304 [quant-ph]},
}

@misc{iannelli_noisy_2021,
	title = {Noisy {Bayesian} optimization for variational quantum eigensolvers},
	url = {http://arxiv.org/abs/2112.00426},
	doi = {10.48550/arXiv.2112.00426},
	urldate = {2025-09-30},
	publisher = {arXiv},
	author = {Iannelli, Giovanni and Jansen, Karl},
	month = dec,
	year = {2021},
	note = {arXiv:2112.00426 [quant-ph]},
}

@Article{Leong2022,
author={Leong, Fong Yew
and Ewe, Wei-Bin
and Koh, Dax Enshan},
title={Variational quantum evolution equation solver},
journal={Scientific Reports},
year={2022},
month={Jun},
day={25},
volume={12},
number={1},
pages={10817},
issn={2045-2322},
doi={10.1038/s41598-022-14906-3},
url={https://doi.org/10.1038/s41598-022-14906-3}
}

@article{Jaderberg2024,
  title = {Let quantum neural networks choose their own frequencies},
  author = {Jaderberg, Ben and Gentile, Antonio A. and Berrada, Youssef Achari and Shishenina, Elvira and Elfving, Vincent E.},
  journal = {Phys. Rev. A},
  volume = {109},
  issue = {4},
  pages = {042421},
  numpages = {10},
  year = {2024},
  month = {Apr},
  publisher = {American Physical Society},
  doi = {10.1103/PhysRevA.109.042421},
  url = {https://link.aps.org/doi/10.1103/PhysRevA.109.042421}
}

@article{Leong2024,
    author = {Leong, Fong Yew and Koh, Dax Enshan and Kong, Jian Feng and Goh, Siong Thye and Khoo, Jun Yong and Ewe, Wei-Bin and Li, Hongying and Thompson, Jayne and Poletti, Dario},
    title = {Solving fractional differential equations on a quantum computer: A variational approach},
    journal = {AVS Quantum Science},
    volume = {6},
    number = {3},
    pages = {033802},
    year = {2024},
    month = {07},
    issn = {2639-0213},
    doi = {10.1116/5.0202971},
    url = {https://doi.org/10.1116/5.0202971},
}

@article{Leong2023,
    author = {Leong, Fong Yew and Koh, Dax Enshan and Ewe, Wei-Bin and Kong, Jian Feng},
    title = {Variational quantum simulation of partial differential equations: applications in colloidal transport},
    journal = {International Journal of Numerical Methods for Heat \& Fluid Flow},
    volume = {33},
    number = {11},
    pages = {3669-3690},
    year = {2023},
    month = {07},
    issn = {0961-5539},
    doi = {10.1108/HFF-05-2023-0265},
    url = {https://doi.org/10.1108/HFF-05-2023-0265},
}

\end{document}